\documentclass[journal]{aa}

\usepackage{graphicx}
\usepackage{amsmath}
\usepackage{amssymb}
\usepackage{verbatim}
\usepackage{array}
\usepackage{txfonts}
\usepackage{natbib}
\usepackage[switch,pagewise]{lineno}
\usepackage{multirow}

\bibpunct{(}{)}{;}{a}{}{,}

\newcommand{\celltspace}{\rule{0pt}{2.8ex}}
\newcommand{\cellbspace}{\rule[-1.4ex]{0pt}{0pt}}

\def\deg{\ensuremath{^\circ}}

\newcommand{\funit}{\,ph\,cm$^{-2}$\,s$^{-1}$}
\newcommand{\msol}{\,$M_{\odot}$}
\newcommand{\rsol}{\,$R_{\odot}$}
\newcommand{\lsol}{\,$L_{\odot}$}
\newcommand{\kms}{\,km\,s$^{-1}$}
\newcommand{\wunit}{\,$M_{\odot}$\,yr$^{-1}$}

\begin{document}

% Title, header, abstract
\title{Particle acceleration and non-thermal emission\\ during the V407 Cygni nova outburst}
\titlerunning{Particle acceleration and non-thermal emission in V407 Cyg}
\author{P.~Martin \and G.~Dubus}
\authorrunning{P.~Martin \and G.~Dubus}
\institute{UJF-Grenoble 1 / CNRS-INSU, Institut de Plan\'etologie et d'Astrophysique de Grenoble (IPAG), UMR 5274, Grenoble, France}
\date{Received: 26 August 2012 / Accepted: 03 January 2013}
\abstract{On March 2010, the symbiotic binary \object{V407 Cyg} erupted as a result of a nova explosion. The event gave rise to a two-week long burst of $\geq$100\,MeV $\gamma$-rays detected by {\em Fermi}/LAT, a unique observation testifying to particle acceleration in the system.}{The outburst can be considered a scaled-down supernova, with short dynamical time scale, and thus can constitute a test case for theories of the origin of Galactic cosmic rays. We aim at determining the properties of the accelerated particles and identifying the origin of the high-energy radiation.}{We developed a model for diffusive shock acceleration and non-thermal emission in \object{V407 Cyg}, complemented by an evaluation of the thermal emission from the shocked plasma. We considered both leptonic and hadronic contributions to the non-thermal processes, and investigated the effect of many binary and nova parameters.}{The $\gamma$-ray emission is mostly of leptonic origin and arises predominantly from inverse-Compton scattering of the nova light. Matching the light curve requires gas accumulation in the vicinity of the white dwarf, as a consequence of wind accretion, while the spectrum imposes particle scattering close to the Bohm limit in the upstream equipartition magnetic field. The nova accelerated protons (respectively electrons) with energies up to $\simeq$300\,GeV (respectively $\simeq$20\,GeV), for a total non-thermal energy $\simeq$10$^{43}$\,erg after two weeks, representing $\simeq$10\% of the initial nova kinetic energy. The electron-to-proton ratio at injection is 6\%.}{The \object{V407 Cyg} eruption can be understood from the same principles that are invoked for particle acceleration in supernova remnants, although without the need for strong magnetic field amplification. The population of novae in symbiotic systems is a negligible source of Galactic cosmic rays, and most likely not a class of TeV-emitters.}
\maketitle

% Introduction
\section{Introduction}
\label{intro}

\indent \object{V407 Cyg} is a binary system belonging to the class of symbiotic binaries, in which a hot compact component is accreting matter from a late-type giant predominantly via a stellar wind. These systems are characterised by composite optical spectra showing absorption features typical of late-type giants and strong nebular emission lines \citep{Belczynski:2000}. \object{V407 Cyg} consists of a white dwarf (WD) and a Mira-type M6 III red giant (RG) with a 745-day pulsation period, which places the system in the group of D-type (dusty) symbiotic stars or symbiotic Miras \citep{Munari:1990}. From the RG pulsation period and Mira period-luminosity relation, the distance to \object{V407 Cyg} is estimated at 2.7\,kpc. Isolated Mira variables with such long pulsation periods generally have thick circumstellar dust envelopes detectable in infrared but causing high obscuration in the optical. In \object{V407 Cyg} (and more generally in symbiotic Miras), however, the influence of the WD through its orbital motion, energetic radiation, and repeated outbursts inhibits dust formation over most of the Mira wind except in a shadow cone behind the RG. As a consequence, \object{V407 Cyg} can be observed in optical, with characteristic obscuration events due to the dust cone. An orbital period of about 40\,yr was inferred from the dust obscuration events, which for a RG mass of 1.0\msol\ and a WD mass of about 1.2\msol\ corresponds to an orbital separation of about 16\,AU \citep{Munari:1990}.\\
\indent \object{V407 Cyg} has been monitored in optical for about 75 years, although irregularly. Apart from the long-term modulation of the Mira pulsed emission by dust obscuration, two episodes of enhanced activity were observed in 1936 and 1998 \citep[the first one being the event that led to the discovery of \object{V407 Cyg}; see][]{Hoffmeister:1949}. On these occasions, the optical emission of the WD rivaled that of the Mira star at its maximum and then slowly declined over the following several years. These flares are typical of all symbiotic binaries and probably result from changes in the accretion rate \citep{Kolotilov:2003}. In 2010, \object{V407 Cyg} was observed on March 10.813 (UT) at V=7, about 5 magnitudes above the maximum of the Mira cycle, a brightness level never reached before in the photometric history of the system \citep{Nishiyama:2010}. Early spectroscopic analyses described the event as a He/N nova expanding in the wind of the Mira companion, with some similarity with the 2006 outburst of \object{RS Oph} \citep{Munari:2010}. Follow-up observations of \object{V407 Cyg} were performed in various bands including radio, infrared, X-rays, and $\gamma$-rays. In particular, the nova explosion was observed in $\gamma$-rays $\geq$100\,MeV by {\em Fermi}/LAT, and it was the first detection of such an object at high energies.\\
\indent In the present paper, we focus on the non-thermal emission of \object{V407 Cyg}, especially that detected by {\em Fermi}/LAT. This radiation testifies to rapid particle acceleration in the nova blast wave as it propagates into the RG surroundings. Whereas in classical novae shock and ejecta are thought to propagate almost freely in a tenuous ambient medium, in symbiotic systems they interact with a denser environment and are thus slowed-down more efficiently as larger quantities of material are swept up and heated up (thermally or non-thermally). This makes \object{V407 Cyg} a short-lived particle accelerator, which in some respect behaves like a scaled-down supernova remnant (SNR) with a life time of just a few weeks or months. The system therefore can constitute a valuable test case for theories of the origin of Galactic cosmic rays. We present a model for diffusive shock acceleration and non-thermal emission in \object{V407 Cyg}, complemented by an evaluation of the thermal emission from the shocked plasma. We considered both leptonic and hadronic contributions to the non-thermal processes, and investigated the effect of many binary and nova parameters. From a comparison to the various observables, we derived estimates for the maximum particle energies, the non-thermal energy budget, the relative contribution of electrons and protons to the emission, and the parameters of the binary system and nova event.

% Observations
\section{Observations}
\label{obs}

\indent In contrast with its previous flares of 1936 and 1998, the outburst experienced by \object{V407 Cyg} in March 2010 was a thermonuclear runaway, the violent burning of a shell of accreted material in degenerate conditions. The high temperatures reached cause the rapid expansion of the shell and eventually its partial ejection from the WD, at velocities of order 1000\kms. Emission line widths provide information about flow velocities after the outburst. H$\alpha$ observations on 13 March revealed a line width consistent with an expansion velocity of the emitting material of 2760 \kms (FWHM, full width at half maximum). Such a broad line originates either from the material ejected in the nova eruption or from the post-shock medium (through charge-exchange), and the inferred velocity at day 2.3 after optical detection is a lower limit but most likely a close estimate of the initial speed of the ejecta. Subsequent line observations revealed that the bulk of the emitting material is progressively slowed down, with an FWHM expansion velocity dropping to 1500\kms\ at day 6.3 and then to 400\kms\ at day 48.2 \citep{Munari:2011}. Yet, the extreme radial velocities of the H$\alpha$ line profile are still about 2500\kms\ even at day 48.3 after optical detection, suggesting that parts of the ejecta were hardly decelerated \citep{Shore:2011}.\\
\indent The outburst in $>$100\,MeV $\gamma$-rays started on the second half of 10 March 2010, the same day as that of the optical maximum (note that the exact start of the nova is uncertain by up to 3 days because the last pre-outburst optical observation was carried out on 8 March). The flare lasted for two weeks and a peak was reached between 13 and 14 March with a flux a factor of 2 higher than that of the initial detection. The average spectral energy distribution over the active period is a power law with a significant exponential cutoff, and a total flux $>$100\,MeV of (4.4 $\pm$0.4) $\times$ 10$^{-7}$\funit. No evidence for spectral variability was found over the active $\gamma$-ray period.\\
\indent In the radio band, measurements at 30\,GHz were made by the OCRA collaboration\footnote{See http://www.jb.man.ac.uk/research/ocra/v407cyg} on several epochs between 24 March 2010 and 3 October 2010 . They revealed an emission at the 25\,mJy level around day 15, rising to a peak flux density of about 45\,mJy at day 40, and then followed by a drop back to about 25\,mJy at day 100. Other observations came from the JVLA Nova Team\footnote{See https://safe.nrao.edu/evla/nova/\#v407cyg}, at several frequencies between 1.4 and 45\,Ghz and covering a period from day 24 to 470. These data show that the radio maximum occurs later and lasts longer at lower frequencies: at 1-2\,Ghz, a flat maximum at 6-7\,mJy is observed over day 50 to 150 approximately. The spectrum of the radio emission flattens progressively over time, from about $F(\nu) \propto \nu^{0.7}$ at day 24 to an almost flat spectrum at day 360.\\
\indent X-ray emission from \object{V407 Cyg} was detected with {\em Swift}/XRT in the 0.3-10\,keV band. Observations started 3 days after the optical maximum and lasted for about 4 months with recurrent visits every few days. Over that period, the emission rose rapidly from about day 10, reached a maximum at day 30 with a luminosity $L_X \simeq 1-2 \times 10^{34}$\,erg\,s$^{-1}$, declined slowly until day 60, and then remained approximately constant at about half the peak luminosity until day 90 \citep{Shore:2011}. The X-ray spectrum consists of two components: a soft blackbody and hard, optically thin emission from a high temperature plasma \citep{Nelson:2012}. The blackbody component is consistent with the emergence of supersoft emission from the nova whereas the hard X-rays trace shock-heated material, as previously seen in \object{RS Oph} \citep{Sokoloski:2006}.\\
\indent In the following section, we describe our model for the nova outburst and accompanying non-thermal processes. The key facts that we sought to reproduce are the early appearance and rapid drop of $\gamma$-ray emission over days 0-14, the level of the thermal emission over days 20-40, and the effective slowing-down of at least a fraction of the shock front at day 40 after outburst. As demonstrated below, thermal and non-thermal constraints should be considered simultaneously because they do not necessarily lead to the same requirements.
\begin{figure}[t]
\begin{center}
\includegraphics[width= 8cm]{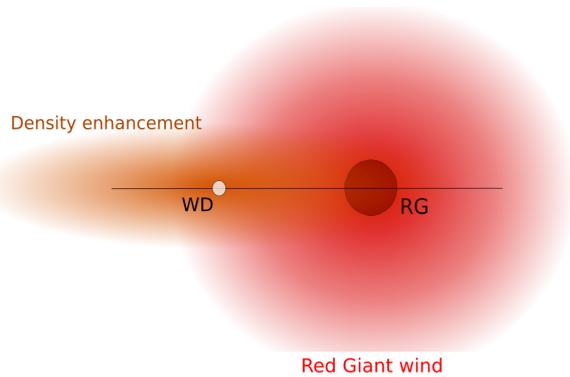} 
\caption{Illustration of the density structure adopted for the medium through which the shock travels. It comprises a spherically-symmetric wind profile about the RG and possibly a Gaussian-shaped matter accumulation around the WD.}
\label{fig_cbm}
\end{center}
\end{figure}

% The model
\section{The model}
\label{model}

% Hydrodynamics
\subsection{Hydrodynamics}
\label{model_hydro}

\indent We assumed that the nova eruption led to the instantaneous ejection of a thin shell of ejecta. This drives a shock wave in the medium, where the sound speed is about a few\kms. The discontinuity associated with the shock front is a fundamental ingredient of the first-order Fermi acceleration process, and our modelling of the nova hydrodynamics therefore focused on the shock position and velocity. The nova shell and associated blast wave were assumed to follow the typical behaviour of a simple non-radiative SNR \citep[see for instance][]{Truelove:1999}. In such a case, the early evolution is usually divided into two stages: first an ejecta-dominated stage (hereafter ED-stage), where the ejecta expand freely and the shock moves at a constant speed, both being largely unaffected by the surrounding medium; second a Sedov-Taylor stage (hereafter ST-stage), where the majority of the ejecta kinetic energy has been transferred to the swept-up ambient gas and the flow evolves adiabatically. For each phase, we adopted the limiting self-similar solutions and neglected the progressive non-self-similar transition in between. Instead, the change from ED-stage to ST-stage was set to happen when the swept-up mass equals the ejecta mass.\\
\indent While in classical novae the shell essentially expands freely into a tenuous circumstellar medium, in \object{V407 Cyg} the ejecta have to expand into a dense environment owing to the thick RG wind. The nova shell can therefore be efficiently slowed down, especially the parts moving in the direction of the RG. The blast wave in \object{V407 Cyg} rapidly becomes aspherical because of the anisotropic matter distribution at the WD position. We therefore adopted a cylindrical geometry with origin at the WD position and symmetry in azimuth, about the WD-RG axis. Any point is characterised by an angle $\theta$ from the symmetry axis and a distance $r$ from the WD (with $\theta=0\deg$ being the direction from the WD to the RG). In our most basic model, the WD is embedded in the RG wind, which is assumed to be spherically-symmetric and to have constant velocity out to large distances. Yet, because of the accretion and orbital motion of the WD, the matter distribution before outburst may be more complex than the simple wind profile, with accumulation of gas in the orbital plane and around the WD \citep[see][for the case of \object{RS Oph}]{Walder:2008}. We therefore also considered a possible circumstellar density enhancement (CDE), although still in the 2D approximation. In the most general case, the density $\rho_{\rm total}$ in the surrounding medium consists of two terms:
\begin{align}
&\rho_{\mathrm{w}}(\theta,r) = \rho_{\mathrm{w,0}} \left( \frac{1}{r^2+d_{\mathrm{orb}}^2-2 r d_{\mathrm{orb}} \cos{\theta}} \right) \label{eq_wind} \\
&\rho_{\mathrm{CDE}}(\theta,r) = \rho_{\mathrm{CDE,0}} \, \exp \left(-\left( \frac{r\sin{\theta}}{b_{\mathrm{CDE}}} \right)^2 -\left( \frac{r\cos{\theta}}{l_{\mathrm{CDE}}} \right)^2 \right) \label{eq_CDE}
\end{align}
where $\rho_{\mathrm{w}}(\theta,r)$ and $\rho_{\mathrm{CDE}}(\theta,r)$ are the wind and CDE density, respectively. The wind profile is defined by its normalisation, $\rho_{\mathrm{w,0}}$, which is obtained from the stellar mass-loss rate $\dot{M}$ and constant wind velocity $V_w$. The CDE profile is defined by its normalisation, $\rho_{\mathrm{CDE,0}}$, and by its characteristic scale lengths, $l_{\mathrm{CDE}}$ and $b_{\mathrm{CDE}}$. The parameter $d_{\mathrm{orb}}$ is the orbital separation. The form of the CDE was taken from a recent study of the X-ray thermal emission from \object{V407 Cyg} (see Sect. \ref{xraysynch_xray}). Modelling of the nova optical light curve also suggests the presence of a large equatorial disc \citep{Hachisu:2012}.\\
\indent The shock wave evolution was computed from the following expressions for the shock velocity $V_{\mathrm{s}}$ in direction $\theta$ and at time $t$:
\begin{align}
V_{\mathrm{s}}(\theta,t) &= V_{\mathrm{ej}} \qquad \qquad \quad \textrm{ for } M_{\mathrm{su}} \leq \frac{M_{\mathrm{ej}}}{4 \pi} \\
V_{\mathrm{s}}(\theta,t) &= \sqrt{\frac{M_{\mathrm{ej}} V_{\mathrm{ej}}^{2}}{4 \pi M_{\mathrm{su}}}} \qquad \, \textrm{ for } M_{\mathrm{su}} >  \frac{M_{\mathrm{ej}}}{4 \pi}
\end{align}
where $M_{\mathrm{ej}}$ and $V_{\mathrm{ej}}$ are the mass and initial velocity of the ejecta. $M_{\mathrm{su}}$ is the swept-up mass per unit solid angle, a quantity that increases with time but at a rate depending on the direction:
\begin{align}
& M_{\mathrm{su}}(\theta,t)= \frac{1}{2 \pi \, {\rm sin}\theta \, {\rm d}\theta} \int_{R_{\mathrm{WD}}}^{R_{\mathrm{s}}} 2 \pi \rho_{\rm total}(\theta,r) \, r^{2} \, {\rm sin}\theta \, {\rm d}\theta \, {\rm d}r \\
& \qquad \textrm{ with } \, R_{\mathrm{s}}(\theta,t)= \int_{0}^{t} V_{\mathrm{s}}(\theta,t) \, {\rm d}t
\end{align}
where $R_{\mathrm{s}}$ is the shock radius and $R_{\mathrm{WD}}$ the WD radius. An implicit approximation in the above equations is that the shock structure has a negligible thickness and can be characterised at any $(\theta,t)$ by a single position and a single velocity. 

% Particle acceleration
\subsection{Particle acceleration}
\label{model_acc}

\indent Our model for particle acceleration is inspired from \citet{Ball:1992} and includes two zones where non-thermal particles are confined: an acceleration zone (zone A) fed by a constant fraction of the freshly swept-up material and where particles undergo diffusive shock acceleration (DSA) and competing energy losses; a cooling zone (zone B) fed by the downstream advection of particles from the acceleration zone and where particles undergo energy losses only. We use a thin-shell approximation and like the hydrodynamic shock structure, both zones are assumed to be coincident with the forward shock. This is a simplification especially for the cooling zone, since advected particles can in principle spread over a certain length in the downstream flow \citep[although that length remains a small fraction of the shock radius, see][]{Schure:2010}. In our model, they remain trapped just behind the forward shock and experience the immediate post-shock physical conditions. In addition, acceleration is assumed to occur at the forward shock only (no contribution from the reverse shock). The calculations are done in the test particle approximation, which means that we neglected the non-linear feedback effects of the non-thermal population on the shock structure and acceleration process.\\
\indent In the acceleration zone, charged particles with sufficient kinetic energy can cross the shock multiple times and are scattered by magnetic perturbations on either side of the forward shock. Because of the velocity discontinuity, particles experience a Lorentz boost at each shock crossing before being almost isotropised in the local fluid frame. The average rate of momentum gain for the DSA process is
\begin{align}
\label{eq_dpdt}
\left( \frac{dp}{dt} \right)_{\mathrm{DSA}} = \left( \frac{q-1}{3q} \right) \frac{V_{\mathrm{s}}^2}{D \left( 1 + q \right)} \, p
\end{align}
where $q$ is the shock compression ratio (which is computed at all $(\theta,t)$, but remains equal to 4 over the majority of the shock surface), $p$ is the particle momentum, and $D$ is the spatial diffusion coefficient. In the most general case, the diffusion coefficient has a dependence on space and momentum coordinates. In our model, we assumed that diffusion properties do not vary over the extent of the accelerator, hence the use of a single coefficient $D$ that applies both upstream and downstream. This implicitly means similar magnetic field conditions on either side of the shock. In particular, the field is assumed to be highly tangled everywhere (because of strong preexising turbulence upstream), so that we can neglect the issue of parallel versus perpendicular diffusion as a function of the propagation angle \citep[which is thought to be important in some objects, e.g. SN1006; see][]{Volk:2003}.\\
\indent The diffusion coefficient is one of the critical parameters of DSA and largely controls the maximum particle energy that can be attained. Particle scattering must be efficient enough to ensure a large momentum gain rate. That it is the case in many SNRs is now supported by both observational evidence and theoretical developments. Magnetic field amplification by cosmic-ray streaming upstream should provide sufficient magnetic turbulence at the relevant scales for efficient resonant scattering \citep[see][and references therein]{Caprioli:2011}. Particle diffusion is then expected to occur close to the Bohm limit in the amplified field, where the pitch angle scattering mean free path is of order of the gyroradius. In our model, we used a diffusion coefficient that is a multiple of the Bohm diffusion coefficient in the upstream non-amplified field:
\begin{align}
\label{eq_diff}
D(p) = \zeta \frac{p \beta c}{3 e B_{\mathrm{s}}} = \zeta D_{\mathrm{Bohm}}
\end{align}
where $e$ is the elementary charge, $B_{\mathrm{s}}$ the upstream non-amplified magnetic field strength just ahead of the shock front, and $\beta=v/c$ where $c$ is the speed of light. The factor $\zeta$, which will be termed diffusion efficiency in the following, actually hides all the details of magnetohydrodynamic wave amplification, advection, and damping. In the most general case, it is time-variable and evolves as the shock speed and upstream medium conditions change. For simplicity, we considered that $\zeta$ is constant and treated it as a free parameter. Overall, with the adopted hypotheses, the momentum gain rate is independent of $p$.\\ 
\indent DSA is accessible to particles that have a large enough gyroradius to probe both sides of the shock. In the thermal leakage scheme, only a small fraction $\sim 10^{-2} - 10^{-4}$ of the downstream thermal pool can enter the acceleration process as suprathermal particles \citep{Volk:2003}. In that case, electrons are almost excluded because of their lower mass, hence lower momentum even for instant post-shock temperature equilibration; far less electrons are thus expected to be accelerated, and those that are have most likely been preaccelerated to relativistic energies by some microphysical mechanism \citep{Morlino:2009}. Estimates of the electron-to-proton ratio from studies of non-thermal emission from SNRs are in the range $\sim 10^{-2} - 10^{-5}$ \citep{Morlino:2009}, while it is $\sim 10^{-2}$ from measurements of the cosmic-ray flux in the local interstellar medium, after propagation in the Galaxy from multiple sources. In the present work, the amount of particles injected into DSA is taken as a constant fraction of the material crossing the shock. We used different injection fractions for protons and electrons, $\eta_{\mathrm{inj,p}}$ and $\eta_{\mathrm{inj,e}}$, and these were treated as free parameters. Particles enter the acceleration process with a fixed momentum, which we chose to be $p_{\mathrm{inj}}= $1\,MeV/c. The exact value has no real importance because of a degeneracy in our model such that the level of non-thermal energy is proportional to the product $\eta_{\mathrm{inj}} p_{\mathrm{inj}}$. We emphasise that constant injection parameters over space and time is a simplification and that they are very likely dependent on the shock conditions \citep{Blasi:2005,Volk:2003}.\\
\indent The non-thermal populations produced by DSA simultaneously undergo momentum losses due to adiabatic expansion of the zones where particles are confined, and radiative losses through a variety of processes. Electrons are subject to synchrotron and relativistic Bremsstrahlung emission in the dense and magnetised shocked material, and inverse-Compton scattering in the intense radiation field of the RG and nova; protons experience inelastic collisions with gas particles in the downstream medium, thereby creating unstable mesons that decay in $\gamma$-rays (see Sect. \ref{model_rad}). The physical conditions used to compute radiative losses at each time are those of the immediate post-shock medium. For zone A, this can be justified by the fact that particles in the acceleration process spend more time downstream than upstream; for zone B, this is a simplification since successive generations of advected particles experience in principle a continuum of conditions as the shocked gas flows away from the shock front.\\
\indent The evolution of the non-thermal particle distribution $N(p)$ in each zone is computed through the following general equation (with subscripts A or B for zone A or B):
\begin{align}
\label{eq_evol}
\frac{\partial N_{A,B}}{\partial t}  &= \frac{\partial}{\partial p} \left( \dot{p}_{A,B} N_{A,B} \right) - \frac{N_{A,B}}{\tau_{A,B}} + Q_{A,B}
\end{align}
where $\dot{p}$ is the momentum gain/loss rate, $\tau$ the characteristic escape time, and $Q$ the source term. In the acceleration zone, these quantities have the following expressions:
\begin{align}
& \dot{p}_A =  \left( \frac{dp}{dt} \right)_{\mathrm{DSA}} + \left( \frac{dp}{dt} \right)_{\mathrm{adiab}} + \left( \frac{dp}{dt} \right)_{\mathrm{rad}} \label{eq_coeffa_1} \\
& \tau_A =  \frac{q \, (1+q) \, D(p)}{V_{\mathrm{s}}^2} \label{eq_coeffa_2} \\
& Q_A = \frac{\eta_{\mathrm{inj}}}{\mu m_p} \frac{d M_{\mathrm{su}}}{dt} \, \delta (p-p_{\mathrm{inj}}) \label{eq_coeffa_3}
\end{align}
where $m_p$ and $\mu$ are the proton mass and the mean molecular mass, respectively. Particle momentum evolves under the competing effects of DSA, adiabatic expansion, and radiative losses (the subscript $rad$ refers to one or a combination of the processes introduced above, depending on whether the particle is a proton or an electron). The escape time is computed as the ratio of the average cycle time by the escape probability. The source term is a Dirac at the momentum $p_{\mathrm{inj}}$. In the cooling zone, the governing equation is:
\begin{align}
\label{eq_coeffb}
& \dot{p}_B = \left( \frac{dp}{dt} \right)_{\mathrm{adiab}} + \left( \frac{dp}{dt} \right)_{\mathrm{rad}} \\
& \tau_B =  \infty \\
& Q_B = \frac{ N_A }{ \tau_A }
\end{align}
The source term is now the escape from zone A and there is no acceleration there. Once in zone B, particles can only accumulate and lose momentum.\\ 
\indent The environmental conditions just ahead of the shock wave are computed at each time from the following assumptions: a spherically-symmetric wind density profile around the RG and possibly a circumstellar density enhancement around the WD, as presented in Sect. \ref{model_hydro}, with a composition of 90\% hydrogen and 10\% helium in particle number; a magnetic field assumed to be in equipartition with the thermal energy density of the upstream material, taken to be highly turbulent and to be at a constant temperature $T_{\mathrm{w}}$; the spherically-symmetric blackbody radiation fields of the RG and of the burning envelope of the WD.

% Radiation
\subsection{Radiation}
\label{model_rad}

\indent The model described above was used to compute the emissions associated with the shock. The main objective of this work is to account for the non-thermal $\gamma$-ray emission detected in the \textit{Fermi}/LAT range, which can result from inverse-Compton scattering, Bremsstrahlung, and hadronic interactions. We also computed the synchrotron emission because radio observations in the $\sim$1-10\,GHz band could be an additional constraint on non-thermal processes. Last, we evaluated the thermal radiation from the shocked material, as X-ray observations can constrain some parameters of the binary system.\\
\indent \textit{Radiation fields:} The RG star light was modelled as a Planck spectral distribution with a temperature of $T_{\mathrm{RG}}$=2600\,K and a radius $R_{\mathrm{RG}}=500\, {\rm R}_\odot$, corresponding to a total luminosity $L_{\mathrm{RG}}=10^4\, {\rm L}_\odot$. With a peak V-band emission nearly 5 magnitudes brighter than the usual level of emission from the system \citep{Munari:2011}, light from the nova eruption can provide a significant contribution to the ambient photon field. An envelope forms rapidly around the WD after nuclear burning is initiated. Peak visual light corresponds roughly to maximum extension of the photosphere, typically $10^{12}-10^{13}$\,cm \citep{Warner:2003}. Thereafter, the envelope shrinks while maintaining a nearly constant bolometric luminosity $L_{\mathrm{nova}}$ close to the Eddington luminosity before mass loss and the end of nuclear burning terminate the process. Detailed models of fast nova eruptions on massive WDs \citep{Kato:1994} show that peak visual light is reached quickly after ignition, at which point the radius of the photosphere around the WD is $R_{\mathrm{nova}}=5\times 10^{12}$ cm and has a temperature $T_{\mathrm{nova}}=10^4$ K for a bolometric luminosity $L_{\mathrm{nova}}=2\times 10^{38} {\rm\ erg\, s}^{-1}=5\times 10^4 {\rm\ L}_{\odot}$. We found that a very good fit to the initial 50 days of the observed V band light curve is obtained by assuming blackbody emission and a photosphere shrinking as $R_{\mathrm{nova}}=5\times 10^{12} \, (t/0.1\, {\rm day})^{-1/2}$\, cm (keeping $L_{\mathrm{nova}}$ constant, hence $T_{\mathrm{nova}} \propto t^{1/4}$), for an adopted distance of 2.7\,kpc. This simple phenomenological approach allows to calculate the nova radiation field at all points and times. The RG and nova luminosities are only a factor 5 different. The nova field therefore dominates during the early evolution as it is closest to the shock. Radiation from the shock-heated material was neglected because its observed luminosity (in X-rays) is a few 10$^{34}\, {\rm erg\, s}^{-1}$, so that the radiation density is no higher than $10^{-4}$ that of the nova for $R_s\geq 5\times 10^{12}$ cm; in addition, its contribution to the scattered spectrum would be suppressed by the Klein-Nishina cutoff of the inverse Compton cross-section.\\
\indent \textit{Non-thermal emission:} For the leptonic radiation processes, synchrotron and inverse-Compton radiation of high-energy electrons were computed using the standard formula for energy loss rate and spectra from \citep{Blumenthal:1970}. As shown below, relativistic Bremsstrahlung is negligible and its contribution to the spectral energy distribution was not computed. The synchrotron emission spectrum was computed with the approximation of a 90\deg\ pitch angle (which overestimates the total emission power by 3/2). For inverse-Compton scattering on the RG light, the transition to the Klein-Nishina regime occurs at a particle energy $\sim$400\,GeV. For the nova light, the transition is at $\sim$100\,GeV at early times, when nova emission is most relevant. As shown below, accelerated electrons will reach energies of a few 10\,GeV at most, so inverse-Compton emission in the {\em Fermi}/LAT range is thus in the Thomson regime. Inverse Compton emission and losses were calculated in the isotropic approximation for most runs, but we also investigated the influence of anisotropy of the seed photon field on the spectrum seen by different observers (see Sect. \ref{gam_run4}), following the method of \citet{Dubus:2008}. In this case, the anisotropic inverse-Compton calculation includes the finite size of the RG or nova photosphere and requires a discretization in azimuth to take into account the change in the scattering geometry. For the hadronic radiation processes, pion decay emission was computed following the parameterization of \citet{Kamae:2006}. We did not use any nuclear enhancement factor to account for the contribution of nuclei heavier than protons \citep[which would increase the emission level by a factor of about 2 for a solar metallicity; see][]{Mori:2009}.\\
\indent \textit{Thermal emission:} Thermal Bremsstrahlung of the material heated by the shock is responsible for the X-ray emission. The typical mean density of the swept-up material downstream of the shock is computed assuming the shell width $\Delta R_s$ is 10\% of the shock radius. This value appears reasonable based on the numerical simulations of \citet{Schure:2010}. The thermal Bremsstrahlung spectrum was calculated following \citet{Rybicki:1986}, using as post-shock density the mean value defined above and as post-shock temperature
\begin{equation}
T_{\mathrm{s}} = \frac{3}{16} \mu m_p V_{\mathrm{s}}^2 \approx 10^8 \left(\frac{V_{\mathrm{s}}}{3000 \, \mathrm{km\, s}^{-1}} \right)^2 \,\mathrm{K}
\end{equation}
or $kT_s\approx 8.6$ keV. The whole approach is very similar to what was done in \citet{Nelson:2012}. Yet, it suffers from various limitations such as the fixed shell width, or the absence of heat conduction and gas mixing. A comparison with comprehensive hydrodynamical simulations of \object{V407 Cyg} revealed these shortcomings (see Sect. \ref{xraysynch_xray}), and showed that such a model can only be used to get a global estimate on the total amount of swept-up material after a few weeks. Still, as we will see, this provides important constraint on the non-thermal processes.\\
\indent \textit{Absorption:} We have taken into account synchrotron self-absorption and thermal free-free absorption {\em within} the shell (again assuming $\Delta R_s/ R_s=0.1$). Absorption within the RG wind is discussed in Sect. \ref{xraysynch_synch}. For simplicity, we have reduced the radiative transfer in the shell to a plane-parallel approximation. Internal absorption is typically small except at low frequencies $\leq 1$\,GHz and therefore affects only the synchrotron emission. The flux after absorption for each position along a given propagation angle is computed as
\begin{equation}
f_\mathrm{abs} = ( f_\mathrm{sync} + f_\mathrm{free} ) \frac{(1-e^{\tau_\mathrm{sync}+\tau_\mathrm{free}})}{\tau_\mathrm{sync}+\tau_\mathrm{free}}
\end{equation}
where the opacities were computed from the local conditions in the shock element following \citet{Rybicki:1986}. Absorption of $\gamma$-rays due to pair production was neglected because the density of X-ray photons from the shock (the only photons that can interact with GeV photons to create pairs) is too low. Assuming that the shocked material emission is all in 1 keV photons, then their density is $n_X \leq 10^7$ ph\,cm$^{-3}$, leading to a pair production opacity $\tau\sim \sigma_T n_X R_s \leq 10^{-6}$. We have also ignored X-ray absorption of the shock emission. The typical column density for the stellar wind integrated down to the surface of the RG is $N_H\sim 10^{23}$ cm$^{-2}$, which leads to absorption below 3\,keV. Thermal Bremsstrahlung emission is harder at the beginning and the column density sampled becomes smaller as the shock progresses outwards such that absorption of the shock emission after 20 days above a keV or so, in the range where we compare our results to the {\em Swift}/XRT observations, can be neglected.

% Time scales
\subsection{Time scales}
\label{model_timescale}

\indent We give below some estimates for the typical time scales of the different processes involved to facilitate the interpretation of the results presented in the following sections. We used as average environmental conditions those found at the WD position in the base case scenario defined in the next section. The particle density, radiation energy densities, and equipartition magnetic field strength upstream of the shock are respectively
\begin{align}
&n_{\mathrm{gas}}= 3 \times 10^{7}\,\textrm{cm}^{-3} \\
&u_{\mathrm{RG}}= 4 \times 10^{-3}\,\textrm{erg\, cm}^{-3} \\  
&u_{\mathrm{nova}}= 20\,\textrm{erg\, cm}^{-3} \\ 
&B_{\mathrm{s}}= 10^{-2}\,\textrm{G} 
\end{align}
where the nova radiation energy density is calculated at the peak assuming the shock is located at the photosphere. The nova radiation field at the shock location decreases rapidly as it moves away (initially $u_{\mathrm{nova}}\propto t^{-2}$), so that the mean $u_{\mathrm{nova}}$ and $u_{\mathrm{RG}}$ typically become comparable within a week. The particle density used in energy loss calculations is the downstream one and is thus a factor of 4 larger than the above value, at about 10$^{8}$\,cm$^{-3}$, because of shock compression.\\
\indent To begin with, the time scale to accelerate a particle to an energy $E$=10\,GeV is 
\begin{equation}
{t}_\mathrm{DSA} \approx 0.1 \, \zeta \left( \frac{E}{10\, \mathrm{GeV}}\right)\left(\frac{0.01\, \mathrm{G}}{B_{\mathrm{s}}}\right)\left(\frac{3000\, \mathrm{km\, s}^{-1}}{V_{\mathrm{s}}}\right)^2 ~\mathrm{d}
\end{equation}
for the typical initial shock velocity and magnetic field intensity experienced by particles over the first days or weeks (depending on propagation angle). At the lowest energies, Coulomb losses could compete with DSA owing to the relatively high densities. Coulomb losses actually dominate over other loss processes below $\sim$100\,MeV and occur on a time scale \citep{Longair:1994}
\begin{equation}
{t}_\mathrm{Coulomb} \approx 45 \, \left(\frac{10^8 \, \mathrm{cm}^{-3}}{n_{\mathrm{gas}}}\right)\left(\frac{E}{100\, \mathrm{MeV}}\right)~\mathrm{d}
\end{equation}
This is much longer than the acceleration time scale, so we have neglected its influence. Similarly, adiabatic decompression due to shell expansion could compete with DSA especially in the early stages. The corresponding time scale for a typical shock radius and velocity is 
\begin{equation}
{t}_\mathrm{adiab} \approx 3 \, \left(\frac{R_{\mathrm{s}}}{5\,\mathrm{AU}}\right) \left(\frac{3000\, \mathrm{km\, s}^{-1}}{V_{\mathrm{s}}}\right) ~\mathrm{d}
\end{equation}
which is more than one order of magnitude above the time scale for acceleration (and even more for lower energies). Particle acceleration can therefore proceed up to higher energies, where it competes with other energy loss processes. Focusing on electrons first, the synchrotron loss time scale is
\begin{equation}
{t}_\mathrm{sync} \approx 5000 \, \left(\frac{0.01\, \mathrm{G}}{B_{\mathrm{s}}}\right)^2\left(\frac{10\, \mathrm{GeV}}{E}\right) ~\mathrm{d}
\end{equation}
In the case of significant magnetic field amplification, the value for $B_{\mathrm{s}}$ would have to be augmented but we will see below that strong amplification of the mean magnetic field is not required. The energy density in stellar radiation is about 1000 times the magnetic energy density, so inverse-Compton can be expected to dominate synchrotron losses and be characterised by a far smaller time scale
\begin{equation}
{t}_\mathrm{IC} \approx 5 \, \left(\frac{4\times10^{-3}\ {\rm erg\,cm}^{-3}}{u}\right) \left(\frac{10\, \mathrm{GeV}}{E}\right) ~\mathrm{d}.
\end{equation}
where inverse-Compton losses were computed in the Thomson approximation. The last process responsible for electron energy loss is relativistic Bremsstrahlung, which occurs on a time scale \citep{Longair:1994}
\begin{equation}
{t}_\mathrm{Br} \approx 300 \, \left(\frac{10^8 \, \mathrm{cm}^{-3}}{n_{\mathrm{gas}}}\right) ~\mathrm{d} 
\end{equation}
which is far longer than that of inverse-Compton. Now looking at protons, energy losses through inelastic collisions with the shocked gas proceed over a characteristic time
\begin{equation}
{t}_\mathrm{pp} \approx 260 \, \left(\frac{10^8 \, \mathrm{cm}^{-3}}{n_{\mathrm{gas}}}\right) ~\mathrm{d} 
\end{equation}
where we have used a threshold for the reaction at 280\,MeV, a constant inelasticity parameter $\kappa_{pp}=0.5$, and a constant cross-section $\sigma_{pp}=30$\,mbarn \citep[which is a reasonable approximation for proton energies below 1\,TeV, see][]{Kelner:2006}.\\
\indent Overall, for the adopted average conditions, we can expect particle acceleration to proceed up to ultra-relativistic energies. For electrons, the initial acceleration is rapidly stuck at 1\,GeV by inverse-Compton losses on the intense nova radiation field at the early shock position; it later continues up to energies of order 100\,GeV, when electron acceleration is limited by inverse-Compton losses on the lower-density radiation fields found at larger distances from the binary. For protons, acceleration could continue up to energies of order 10\,TeV, but we will see below that it stops before as a result of shock slowing-down. Electrons will therefore very likely be loss-limited and protons age-limited. In terms of radiation in the 100\,MeV-1\,GeV range (which results from 10\,GeV particles, either through inverse-Compton or hadronic interactions), non-thermal electrons are at least $\sim$100 times more efficient in radiating away their energy than non-thermal protons. Yet, if the latter are at least 100 times more numerous than the former, they will dominate the $\gamma$-ray emission.\\
\indent For the thermal emission, the cooling time scale of the shock-heated wind is
\begin{equation}
{t}_\mathrm{ff}\approx 170 \, \left(\frac{T_{\mathrm{s}}}{10^8\, \mathrm{K}}\right)^{1/2}  \left(\frac{10^8 \, \mathrm{cm}^{-3}}{n_{\mathrm{gas}}}\right) ~\mathrm{day}
\end{equation}
using a Gaunt factor equal to 1, and $T_{\mathrm{s}}$ as defined previously. The cooling time scale of the shocked material is long, which justifies the assumption of Sedov-Taylor expansion in the first couple of months, provided most of the shock kinetic energy is indeed channeled into heating up the gas thermally \citep[see also][]{Nelson:2012}.

% Computation
\subsection{Computation}
\label{model_num}

\indent The evolution of the whole system was computed for a grid of propagation angles over successive time steps, using an angular step of $\Delta \theta=\pi/100$\,rad and a time step of $\Delta t=0.1$\,d. The non-thermal particle populations were followed on a momentum grid running from $p_{min}=10^{-1}$ to $p_{max}=10^{8}$ with a logarithmic step of 0.01 (in units of MeV/$c$).\\ 
\indent For a given set of parameters, the evolution is computed along each direction $\theta$, starting at the position of the WD surface with the initial shock velocity. At each time step for a given propagation angle: the shock properties (radius, velocity, swept-up mass, dynamical stage) for the current shock element are updated; the environmental conditions are determined (gas and radiation density, magnetic field strength); the non-thermal particle distributions are computed according to Eq. \ref{eq_evol}, first in zone A and then in zone B. All along, we check that the current shock element does not crash into the RG by a simple criterion on position. If that is the case, the calculation is stopped for the current direction and the associated particle population is removed from the total. We did not handle the interaction of the shock with the RG surface or the convergence of the shock on the rear side of the star, so the shadow cone behind the RG is a dead zone in our model. Once the hydrodynamical and non-thermal evolution is computed, the corresponding radiative outputs are determined in a post-processing stage.\\
\indent For a given layout of the \object{V407 Cyg} binary system, there are three free parameters: the diffusion efficiency, and the injection fractions for protons and electrons. Since we neglected the non-linear aspects of the acceleration process, the calculation is linear and the injection fractions are just normalisation factors, tuned to match the $\gamma$-ray observations. In any case, they are constrained by energetic considerations: the energy channelled into non-thermal components should not exceed a fraction of the total nova kinetic energy. As we will see, the diffusion efficiency provides control on the maximum particle energy, and thus on the $\gamma$-ray spectral maximum. The value eventually required for this parameter provides information about the particle scattering regime and possible magnetic field amplification.\\
\indent We present below the outcome of three different calculations, in order to emphasise how thermal and non-thermal emissions depend on the system properties. The first calculation is based on typical parameters for the \object{V407 Cyg} system, quite similar to those adopted in \citet{Abdo:2010c}, with a surrounding medium consisting only of the RG wind. This base case scenario will be useful in illustrating the main features of particle acceleration in such a system but cannot account for several observations. A second calculation is then presented, where the binary system and acceleration parameters were optimised in order to reduce the discrepancy with the experimental data, but still keeping a wind density profile for the ambient medium. This scenario also has major shortcomings and thus calls for improvement of the model. In a last calculation, a density enhancement was added to the simple wind profile and yielded a far better fit to most observations. These calculations will be referred to as Run 1, Run 2, Run 3, respectively, and their corresponding sets of parameters are summarized in Table \ref{tab_run}. In terms of particle acceleration, we aimed at accounting for the {\em Fermi}/LAT data with the smallest amount of non-thermal energy, in order to remain close to the limit of test particle approximation, and with the smallest electron-to-proton ratio at injection $K_{\mathrm{ep}}=\eta_{\mathrm{inj,e}} / \eta_{\mathrm{inj,p}}$, in order to be in agreement with the values quoted in Sect. \ref{model_acc}. In the base case scenario, for illustration, particle acceleration was assumed to occur in the Bohm limit in the non-amplified equipartition magnetic field (the field strength is already quite high, typically about $10^4$ times the local interstellar value), and the electron-to-proton ratio was set to 1\%. 
\begin{table}[t]
\begin{center}
\caption{Summary of the different calculations presented in the paper.}
\begin{tabular}{|c|l|l|}
\hline
\celltspace Run & Binary system parameters & Acceleration parameters \cellbspace \\
\hline 
\celltspace \multirow{3}{*}{1} & $M_{\mathrm{ej}}$= 10$^{-6}$\msol & $\eta_{\mathrm{inj,p}}=3 \times 10^{-3}$ \\
& $V_{\mathrm{ej}}$= 3000\kms & $\eta_{\mathrm{inj,e}}=3 \times 10^{-5}$ \\
& $d_{\mathrm{orb}}$= 10\,AU & $K_{\mathrm{ep}} = 0.01$ \\
& $\dot{M}_{\mathrm{RG}}$= 5 $\times$ 10$^{-7}$\wunit & $\zeta$=1 \\
& $R_{\mathrm{WD}}$= 0.01\rsol &  \\
& $R_{\mathrm{RG}}$= 500\rsol &  \\
& $L_{\mathrm{RG}}$= 10000\lsol &  \\
& $V_w$= 20\kms &  \\
& $T_w$= 700\,K &  \\
& $\rho_{\mathrm{CDE}}=0$ (no CDE) & \cellbspace \\
\hline
\celltspace \multirow{3}{*}{2} & $M_{\mathrm{ej}}$= 2 $\times$ 10$^{-6}$\msol & $\eta_{\mathrm{inj,p}}=6 \times 10^{-3}$ \\
& $\dot{M}_{\mathrm{RG}}$= 10$^{-7}$\wunit & $\eta_{\mathrm{inj,e}}=6 \times 10^{-4}$ \\
& $d_{\mathrm{orb}}$= 6\,AU & $K_{\mathrm{ep}} = 0.1$ \\
& $\rho_{\mathrm{CDE}}=0$ (no CDE) & $\zeta$=1 \cellbspace \\
\hline
\celltspace \multirow{3}{*}{3} & $M_{\mathrm{ej}}$= 2 $\times$ 10$^{-6}$\msol & $\eta_{\mathrm{inj,p}}=5 \times 10^{-3}$ \\
& $\dot{M}_{\mathrm{RG}}$= 5 $\times$ 10$^{-8}$\wunit & $\eta_{\mathrm{inj,e}}=3 \times 10^{-4}$  \\
& $d_{\mathrm{orb}}$= 10\,AU & $K_{\mathrm{ep}} = 0.06$ \\
& $\rho_{\mathrm{CDE}}=10^{8}$\,cm$^{-3}$ & $\zeta$=3  \\
& $l_{\mathrm{CDE}}=10$\,AU, $b_{\mathrm{CDE}}=5$\,AU & \cellbspace \\
\hline
\end{tabular}
\label{tab_run}
\end{center}
Note to the table: CDE stands for Circumstellar Density Enhancement (see Sect. \ref{model_hydro}). For Run 2 and Run 3, the parameters not listed have the same values as in Run 1.
\end{table}
\begin{figure*}[!ht]
\begin{center}
\begin{tabular}{cc}
\includegraphics[width= 8cm]{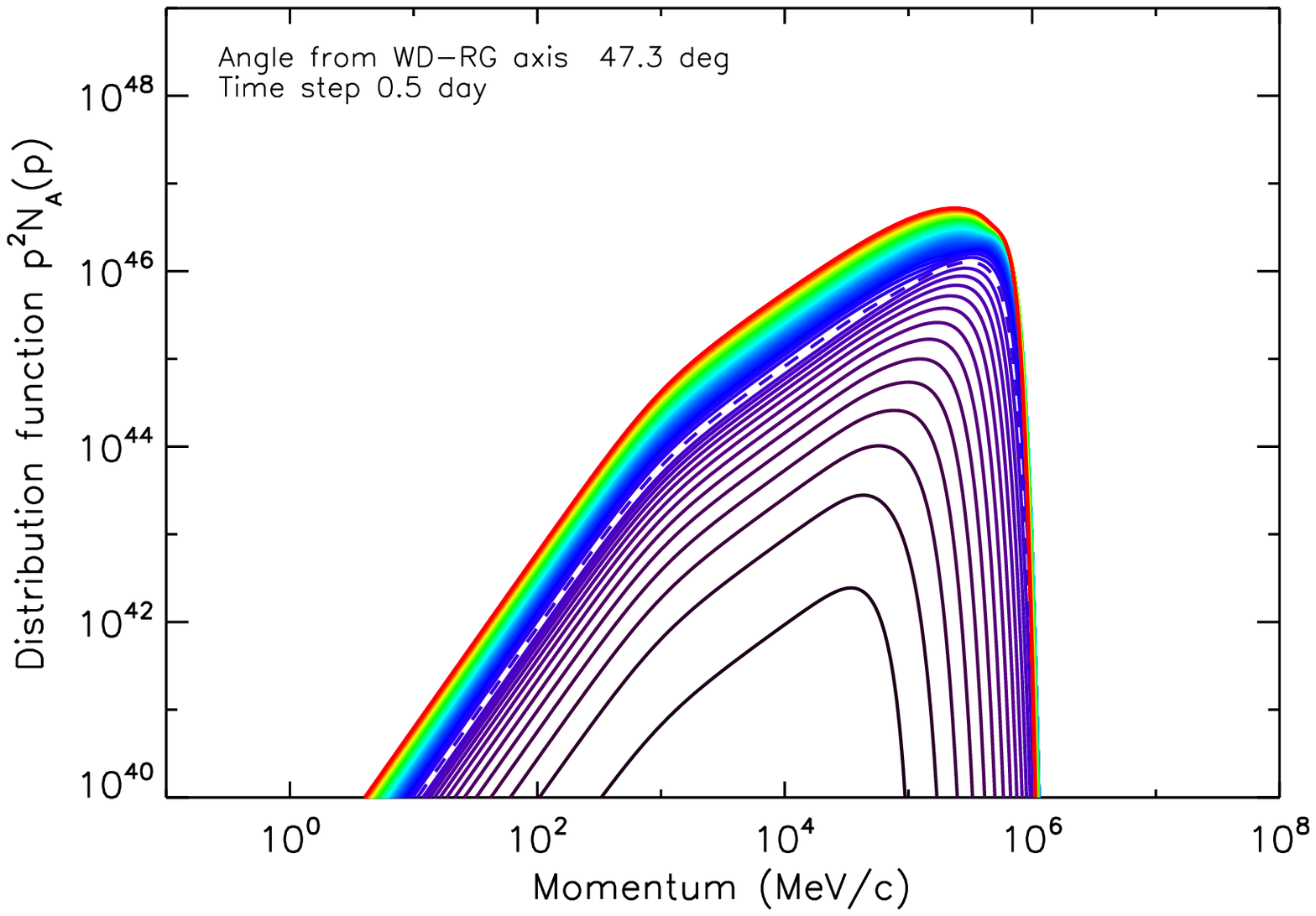} &  \includegraphics[width= 8cm]{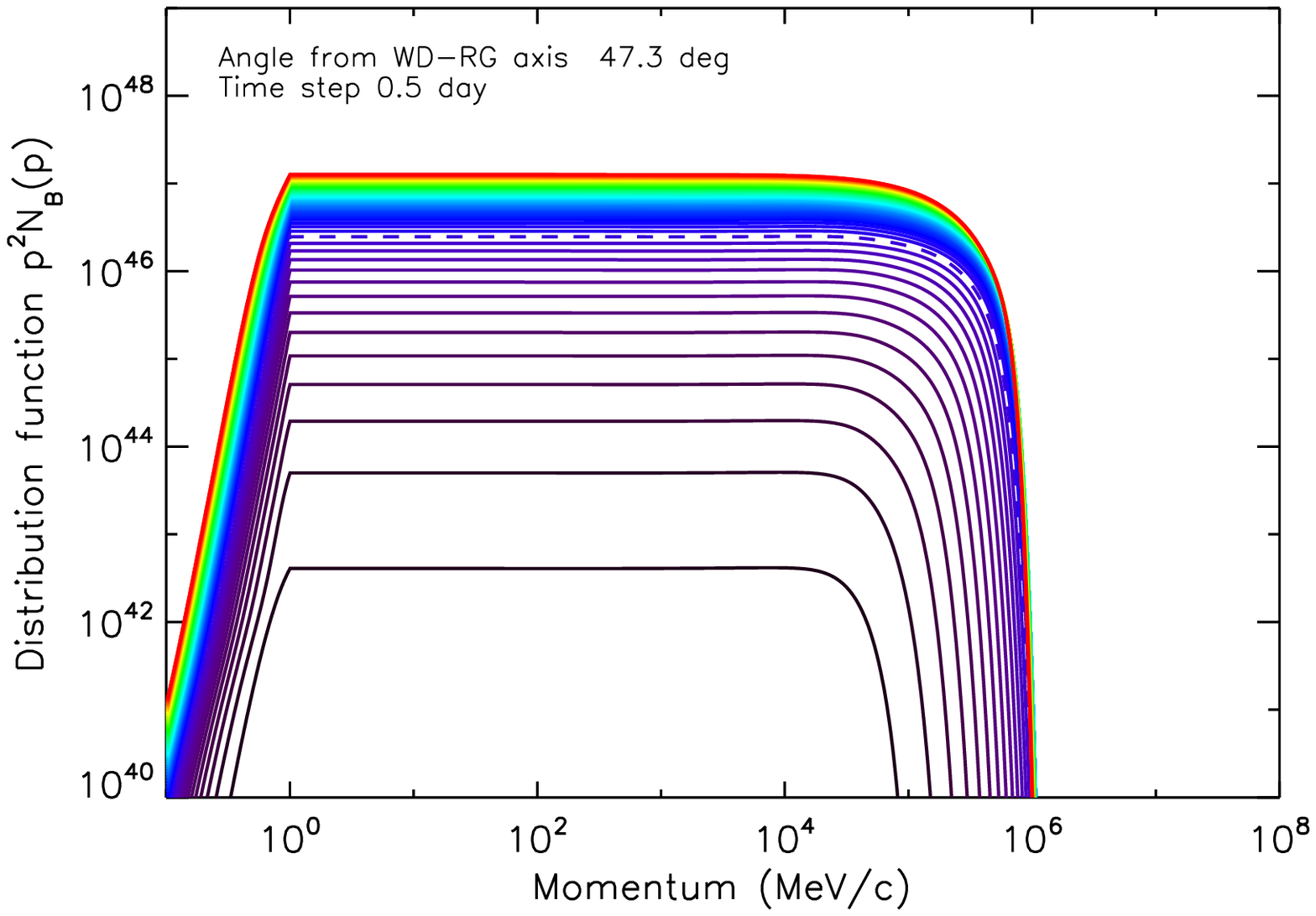} \\
\includegraphics[width= 8cm]{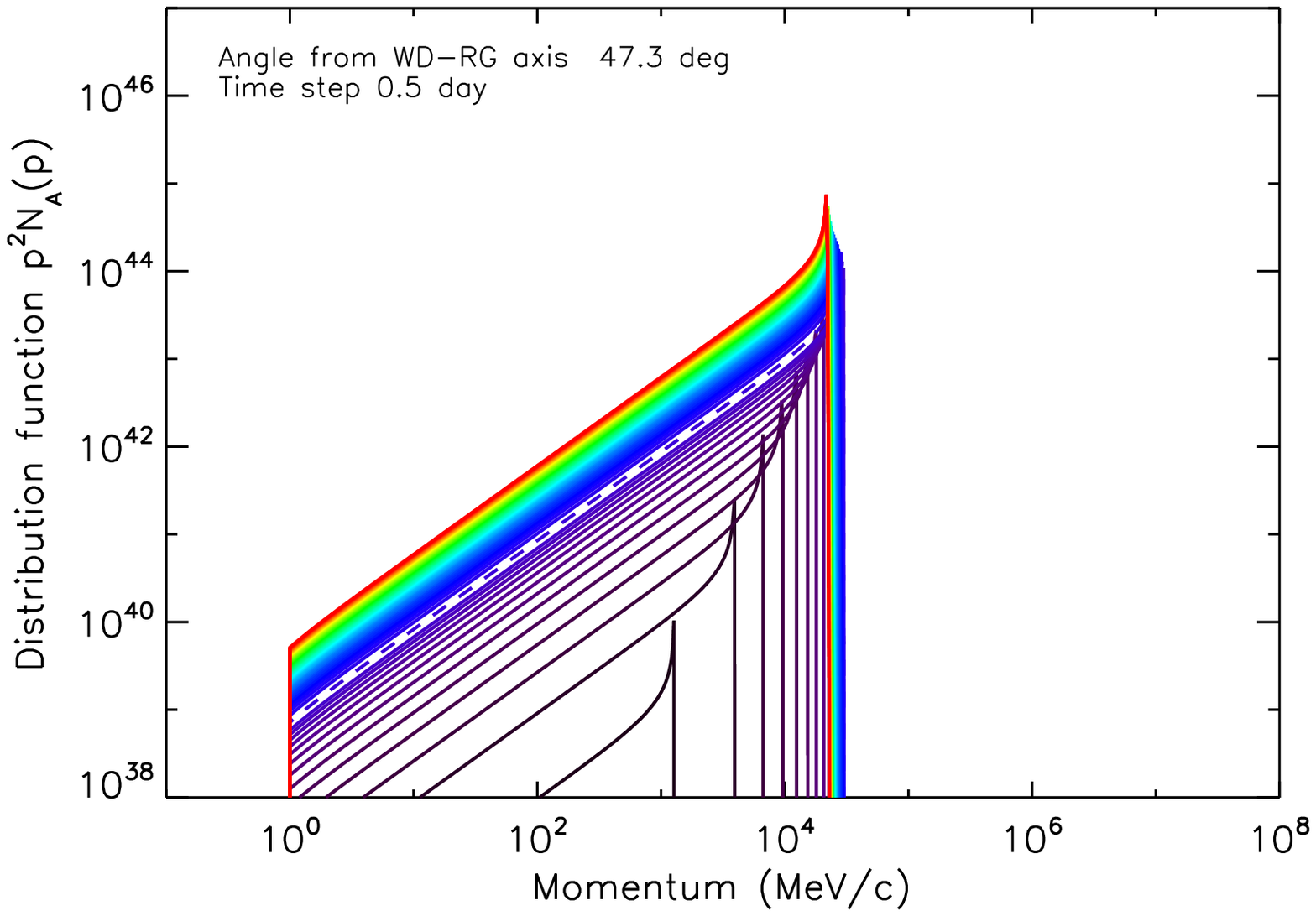} &  \includegraphics[width= 8cm]{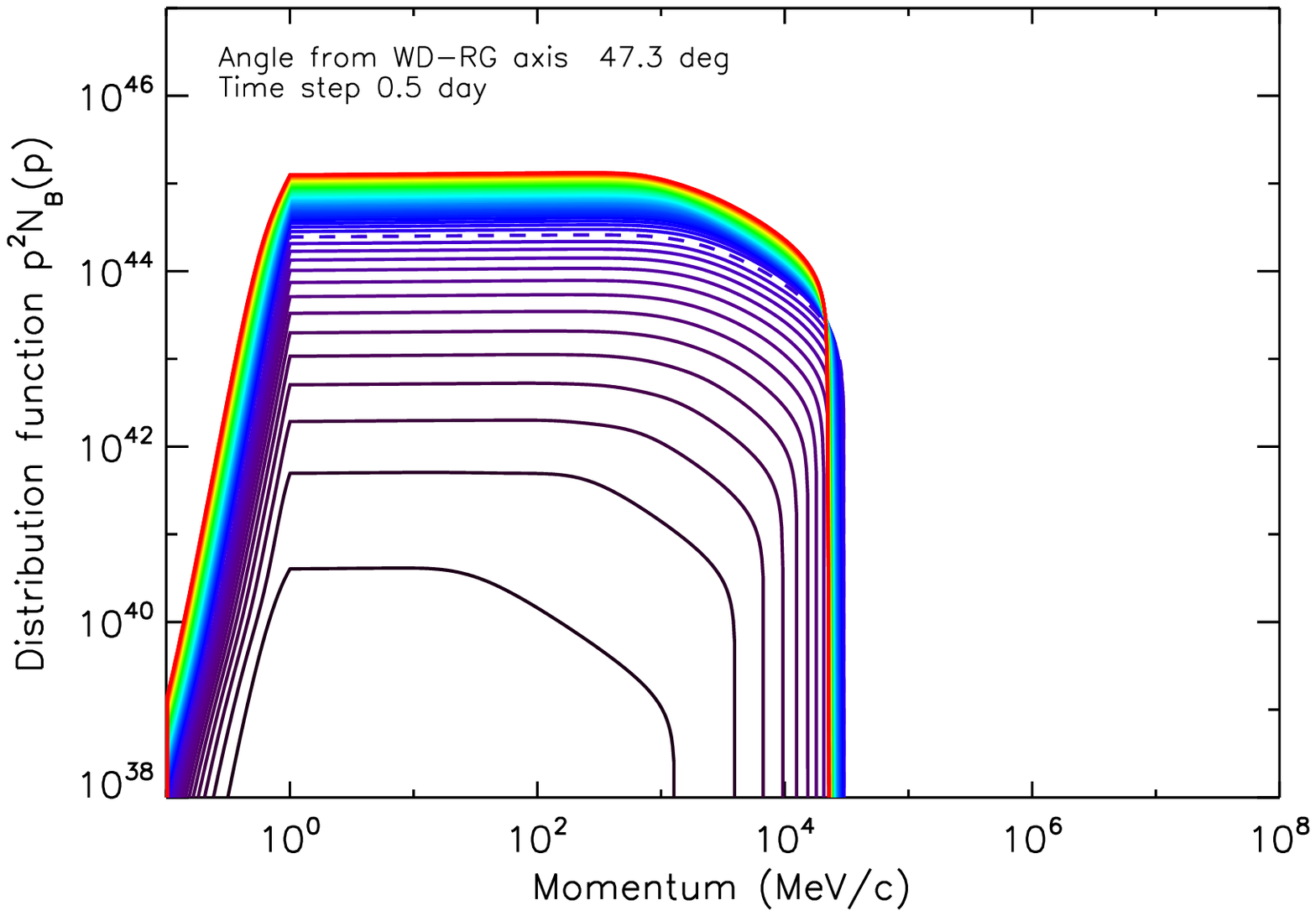}
\end{tabular}
\caption{Time evolution of the proton and electron distributions (top and bottom row, respectively) in the acceleration zone and cooling zone (left and right column, respectively). The results correspond to the base case scenario with energy losses, a propagation angle of 47.3\deg\ from the WD-RG axis, and injection fractions of $\eta_{\mathrm{inj,p}}=10^{-3}$ and $\eta_{\mathrm{inj,e}}=10^{-5}$. Curves from black to red show successive time steps of 0.5\,day. The dashed curve in each plot marks the ED to ST transition.}
\label{fig_distrib_basecase_angle}
\end{center}
\end{figure*}

% Non-thermal particle populations
\section{Non-thermal particle populations}
\label{pop}

\indent In this section, we focus on the non-thermal particle populations that can be produced by \object{V407 Cyg} in the base case scenario (as defined in Sect. \ref{model_num}). We present their evolution in time and space over the first 40 days after the nova outburst. We first discuss the properties of the particle distributions especially their maximum energy and the impact of energy losses, then the effects of the asphericity of the phenomenon on particle acceleration, and last the global energetics.

% Distribution properties
\subsection{Distribution properties}
\label{pop_distrib}

\indent In Fig. \ref{fig_distrib_basecase_angle}, the time evolution the non-thermal particle populations in the acceleration and cooling zones is plotted, for a propagation angle of about 45\deg. The distributions obtained for other directions are very similar, except the ones for low angles (propagation towards the RG), whose evolution is  prematurely halted by the collision with the star.\\
\indent We first discuss the case of accelerated protons, which are almost only exposed to acceleration and downstream advection, hadronic interactions and adiabatic losses being not efficient enough. Therefore, they approximately illustrate the case of acceleration without energy losses. In the acceleration zone, relativistic particles soon follow a power-law distribution in momentum with a slope $s=-1$ and a cutoff at an increasing momentum. In the cooling zone, relativistic particles also follow a power-law distribution with cutoff, but steeper with $s=-2$ because of the energy dependence of the advection time, which favours the escape of lower-energy particles (see Eq. \ref{eq_coeffa_2})\footnote{The proton distributions for zone A exhibit a break at about 1\,GeV/$c$ that marks the transition to the relativistic regime. This feature arises because of the $\beta$ factor in the diffusion coefficient $D(p)$, which implies a very efficient acceleration of mildly-relativistic particles. In the same time, this means an easier downstream advection of the same particles, so both effects compensate and the break does not appear in the proton distributions for zone B (see Eqs. \ref{eq_diff} and \ref{eq_coeffa_2}).}. The cutoff in both distributions is set by the finite acceleration time, hence its progressive shift to higher and higher momenta. This is especially true for acceleration during the ED-stage, where the acceleration rate remains almost constant because of a constant shock velocity. The progression to higher energies seems to stall because the acceleration rate is independent on the particle momentum (see \ref{model_acc}), and so a given logarithmic increase in energy takes longer to be achieved at higher momenta. In addition, for those parts of the shock front that enter in ST-stage, the acceleration rate decreases because of shock slowing-down, and the cutoff momentum hardly evolves anymore. It is, however, during the ST-stage that most particles are accelerated. Overall, non-thermal protons are almost age-limited. In the base case scenario, the maximum particle energy is slightly short of 100\,GeV after just 1\,day, and it increases by about one order of magnitude to almost reach 1\,TeV when the shock starts to slow down. After that, the maximum particle energy remains almost constant up to at least day 40.\\
\indent The situation is somewhat different for accelerated electrons. Their distribution is strongly modified by inverse-Compton losses, as illustrated by Fig. \ref{fig_distrib_basecase_angle}. In the acceleration zone, electrons are initially stuck at GeV energies because of the very strong inverse-Compton losses on the nova light (if they were exposed to the RG light only, they would move up to about 50\,GeV in just one day). The competition of acceleration and inverse-Compton results in a pile-up at the high end of the distribution. As the shock moves away from the nova photosphere, the inverse-Compton loss rate drops and the particles can be pushed to higher energies, eventually reaching 30\,GeV at the end of the ED-stage. In the ST-stage, the acceleration rate decreases and inverse-Compton losses prevail, which causes a progressive shift of the pile-up to lower energies. This behaviour is reflected by the distribution in the cooling zone, which in addition is altered at the high end by inverse-Compton losses and exhibits a characteristic steepening from $s=-2$ to $s=-3$.

% Effect of asphericity
\subsection{Effect of asphericity}
\label{pop_aspher}

\begin{figure}[t]
\begin{center}
\includegraphics[width= 8cm]{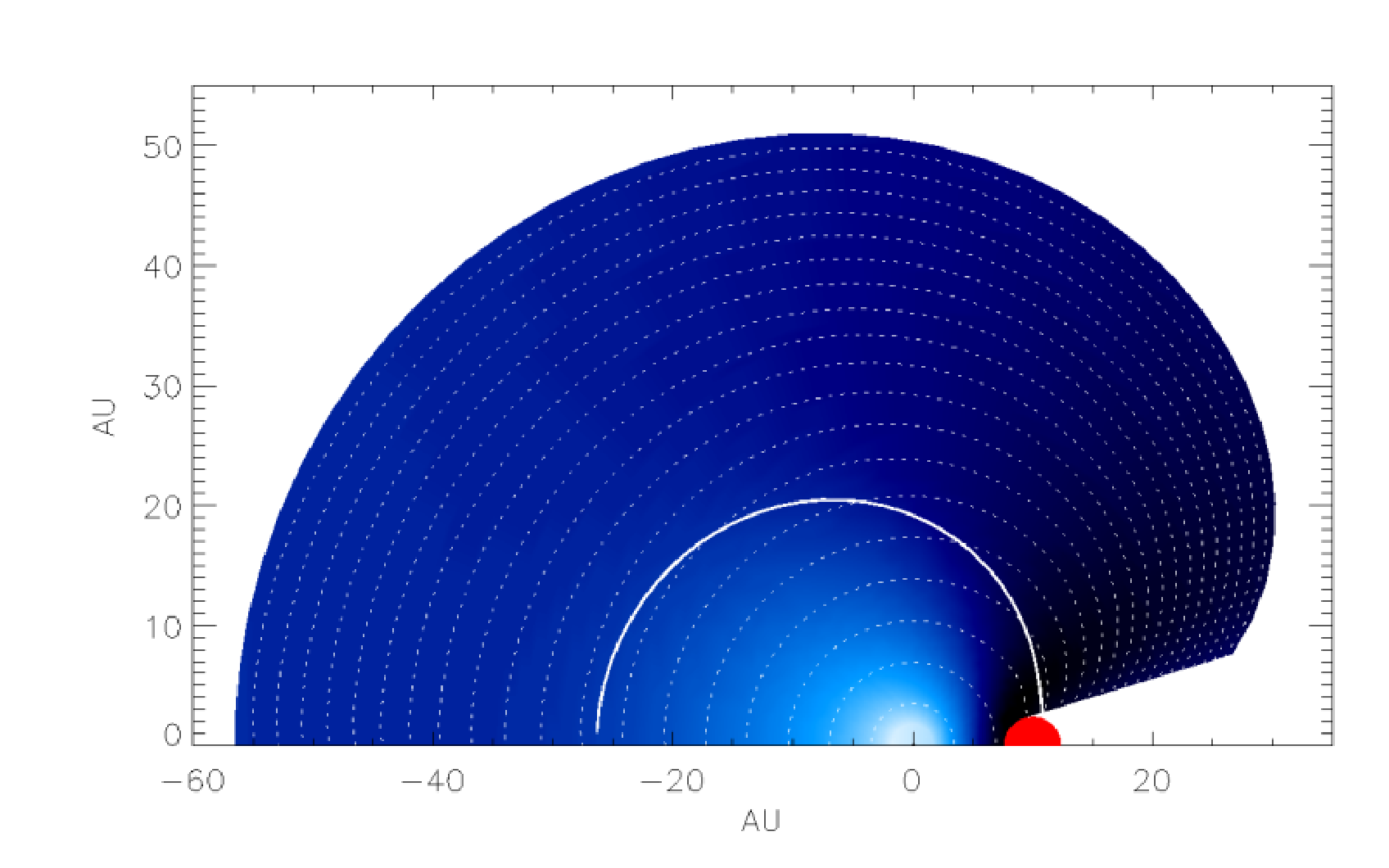}
\includegraphics[width= 8cm]{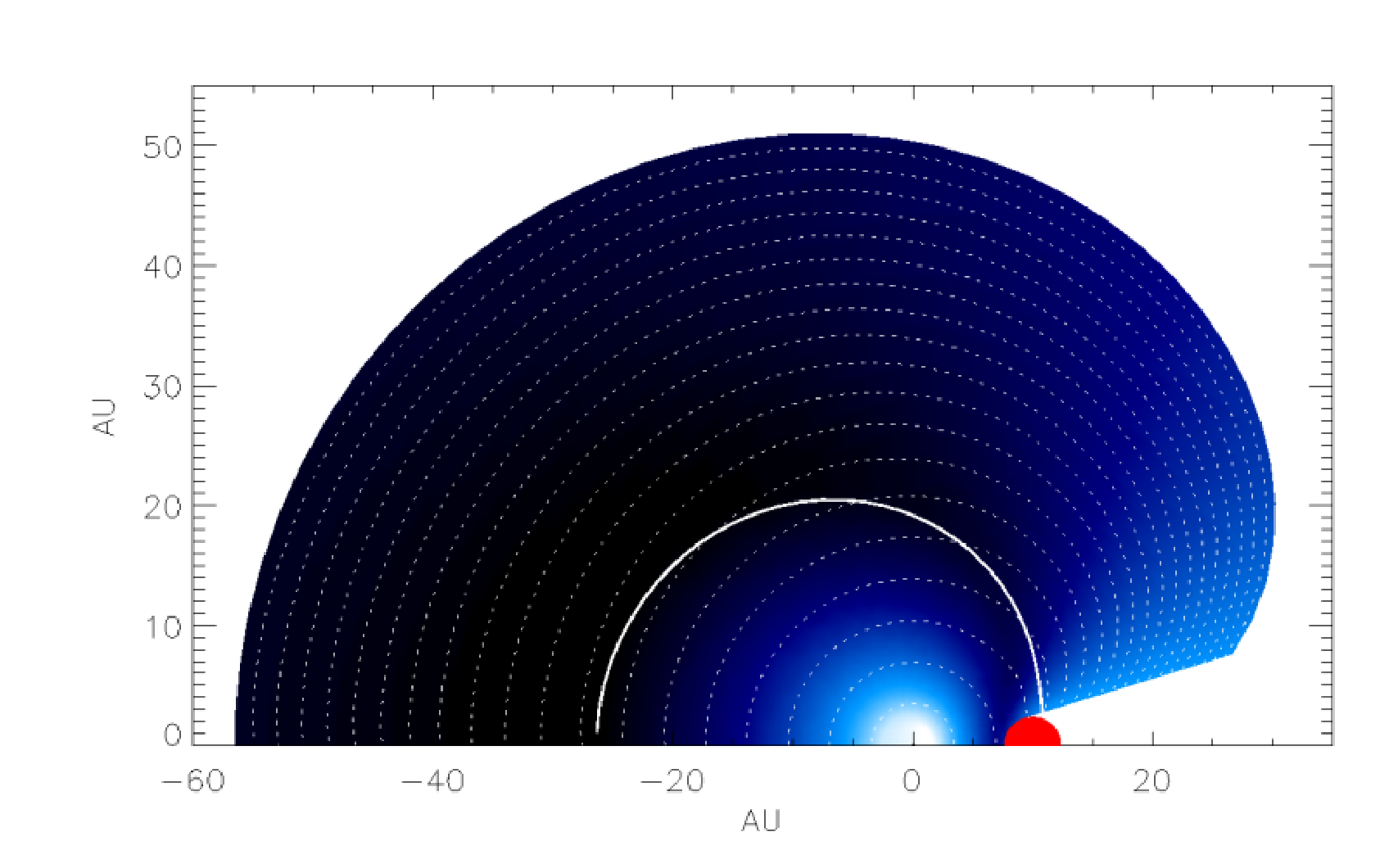}
\caption{Spatial maps of the maximum energy attained by protons (top) and electrons (bottom) in the base case scenario. The color scaling is linear between the highest value (in black) and values that are less than 10 times smaller (in white). Note that the maxima are different for protons and electrons (see text). The WD is in ($x$,$y$)=(0,0) and the RG in ($x$,$y$)=(10,0). Dotted white lines mark successive positions of the shock front with a 2-day time step, and the solid white line marks the ED-ST transition.}
\label{fig_emax_map}
\end{center}
\end{figure}
\begin{figure}[t]
\begin{center}
\includegraphics[width= 8cm]{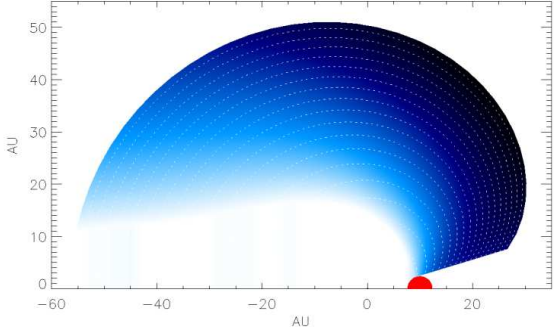}
\caption{Spatial map of the non-thermal energy stored in accelerated protons in the base case scenario (the plot for electrons is quite similar). Color coding and graph elements are described in Fig. \ref{fig_emax_map}.}
\label{fig_ntnrj_map}
\end{center}
\end{figure}

\indent The aspherical shock wave and anisotropic binary environment have an impact on accelerated proton and electron energies and on how the non-thermal energy is distributed over the shock front. This is illustrated in Figs. \ref{fig_emax_map} and \ref{fig_ntnrj_map}.\\
\indent For protons, there is a mild dependence of the maximum particle energy on the direction of shock propagation. The transition from ED to ST-stage occurs later with increasing propagation angle, but this has a limited impact because of the equipartition assumption for the magnetic field: a lower average density along a given propagation direction ensures a longer ED-stage, but implies a smaller magnetic field; eventually, a lower acceleration rate operates for a longer time, giving about the same maximum energy. As a result, the maximum proton energies are quite uniform after a few days and remain almost unchanged subsequently. The highest particle energies are reached for propagation towards the densest regions because of stronger equipartition magnetic fields.\\
\indent For electrons, which are strongly affected by energy losses, the picture is different. For shock elements propagating towards a denser radiation field, inverse-Compton losses limit the maximum particle energy; the highest energies are therefore reached for propagation away from the RG. As a result, the maximum electron energy is achieved over a different domain of the shock trajectory and it is less uniform at a given time than for protons. Another difference is that the maximum energies will decrease past the ED-ST transition whatever the propagation direction, in contrast to protons.\\
\indent Because of the relatively limited variation in maximum particle energy over the shock surface, especially for protons, the non-thermal energy distribution over the shock front is essentially set by the number of accelerated particles, hence the amount of swept-us mass. For a blast wave propagating in a RG wind only, shock elements moving along angles between about 10\deg\ and 90\deg\ are those that dominate the production of non-thermal particles (see Fig. \ref{fig_ntnrj_map}). 

% Global energetics
\subsection{Global energetics}
\label{pop_nrj}

\indent Particle acceleration channels a fraction of the nova kinetic energy to non-thermal components. This energy first goes into non-thermal particles, which transfer it partially to radiation and magnetic fields. In the following, we will define at each time the fraction of the nova energy that is in the form of non-thermal particles as \textit{acceleration efficiency}, and refer to the fraction of non-thermal particle energy that has gone into non-thermal radiation as \textit{radiation efficiency}. Then, the \textit{non-thermal efficiency} is defined as the fraction of the nova kinetic energy channelled into non-thermal components, particles or radiation. In the energy budget, we neglect the magnetic energy drained from accelerated particles by wave amplification processes \citep[this can extract $\leq$10\% of the incoming flow energy for parallel, non-relativistic shocks with Mach number $M_A > 30$;][]{Gargate:2012}.\\
\indent As emphasised previously, the non-linear aspects of the particle acceleration process were not considered in our model. The main consequence is that the energy and flux levels of non-thermal components are proportional to the injection fractions (for a given injection momentum). For the values $\eta_{\mathrm{inj,p}}=3 \times 10^{-3}$ and $\eta_{\mathrm{inj,e}}=3 \times 10^{-5}$ required to match the observed $\gamma$-ray flux (see below), the acceleration efficiencies at day 40 for protons and electrons are about 100\% and 2\%, respectively\footnote{The acceleration efficiency for electrons is not 100 times less than the acceleration efficiency for protons, as could have been expected from the ratio of the injection fractions, because at low energies $<$10\,GeV electrons gain more kinetic energy than protons for the same momentum gain.}. Such injection fractions in the base case scenario therefore seem to be excluded by energetic considerations. As we will see below, however, the correspondence between injection fraction and acceleration efficiency depends on the environmental conditions.\\
\indent Last, it is interesting to compare the content of both zones. The cooling zone contains more particles and energy than the acceleration zone, but the situation is different for electrons and protons. After a few days, there is about 4 times (respectively 30 times) more non-thermal energy in the cooling zone than in the acceleration zone for protons (respectively electrons). This difference is not uniform across the energy range due to the different spectral distributions, as can be seen from Fig. \ref{fig_distrib_basecase_angle}. The cooling zone contains about a few 100 times more $\sim$1\,GeV particles than the acceleration zone, but this contrast diminishes with energy and the contents become comparable at $\sim$100\,GeV. This is so because higher-energy particles take longer to be produced and have a longer escape time. And since protons can reach energies that are 10-20 times higher than that of electrons, they can store more energy in the acceleration zone than electrons.
\begin{figure}[t]
\begin{center}
\includegraphics[width= 8cm]{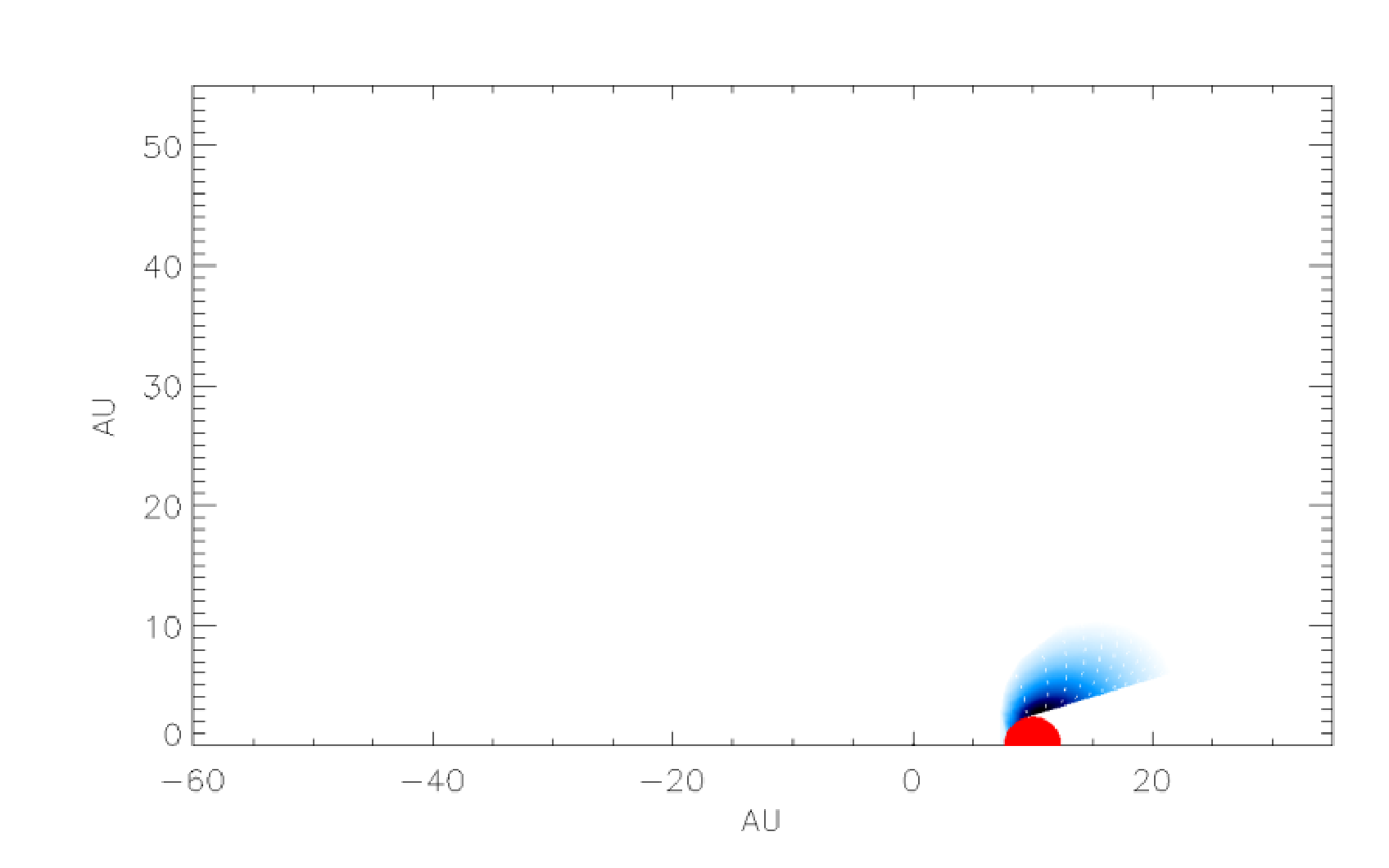} 
\includegraphics[width= 8cm]{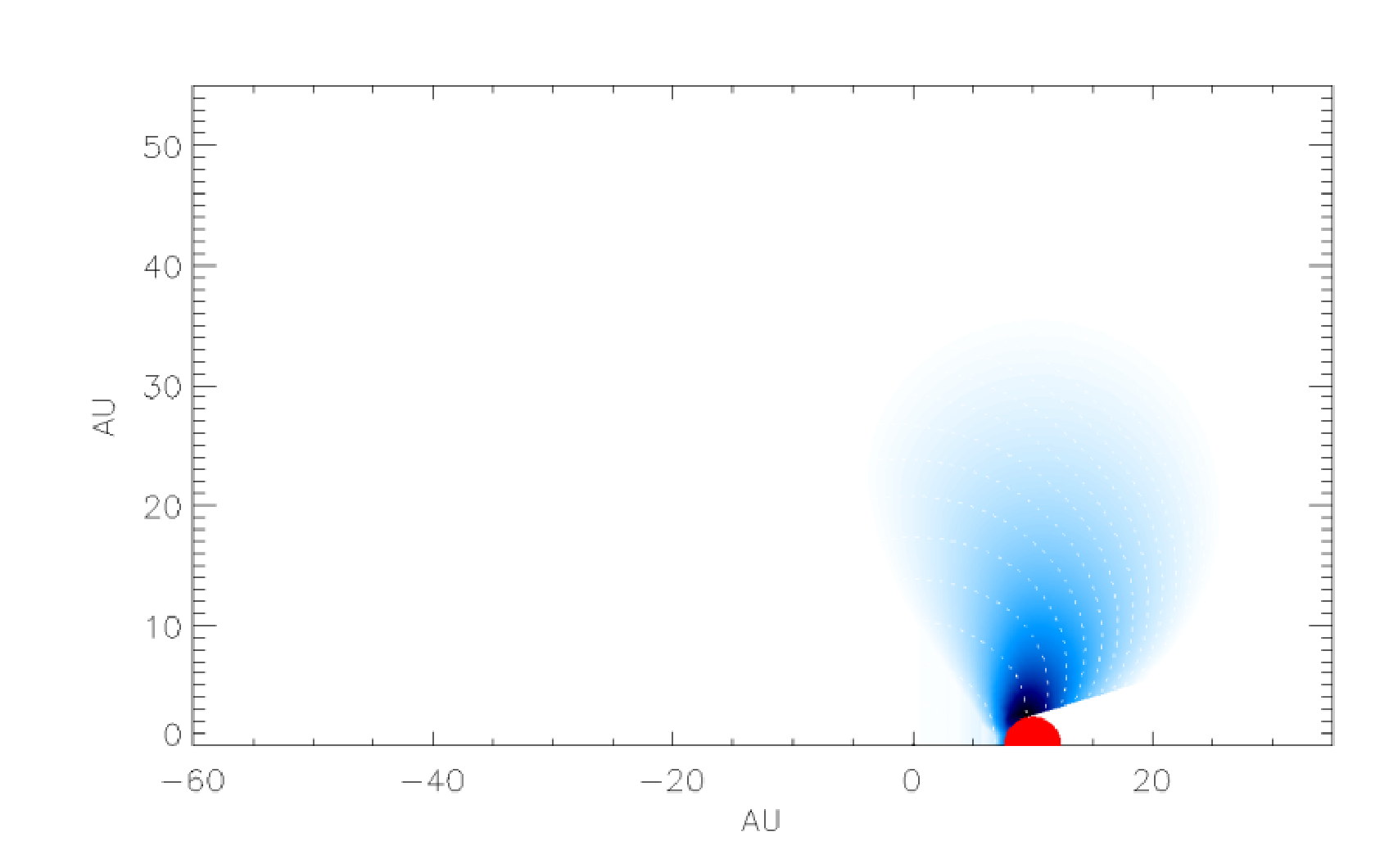}
\caption{Spatial maps of the $>$100\,MeV $\gamma$-ray emission from pion decay (top) and inverse-Compton (bottom) in the base case scenario. Color coding and graph elements are described in Fig. \ref{fig_emax_map}.}
\label{fig_fgamma_map}
\end{center}
\end{figure}

% Gamma-ray emission
\section{Gamma-ray emission}
\label{gam}

\indent In this section, we discuss the $\gamma$-ray emission associated with the \object{V407 Cyg} outburst. We first show the results obtained in the base case scenario, where the shock propagates in the RG wind with typical parameters for the binary system and acceleration process. This illustrates several properties of the high-energy emission from such a system but, as we will see, it cannot account for the observations of \object{V407 Cyg}. We then present an attempt to optimize the binary system and acceleration parameters in order to reduce the discrepancy with the experimental data, and from that conclude that the observations cannot be reproduced for a shock propagating in a wind only. In a third part, we improve our model by considering a more realistic ambient medium that includes a density enhancement on top of the wind profile. 

% Shock propagating in a wind: base case scenario
\subsection{Shock propagating in a wind: base case scenario}
\label{gam_run1}

\begin{figure}[!ht]
\begin{center}
\includegraphics[width= \columnwidth]{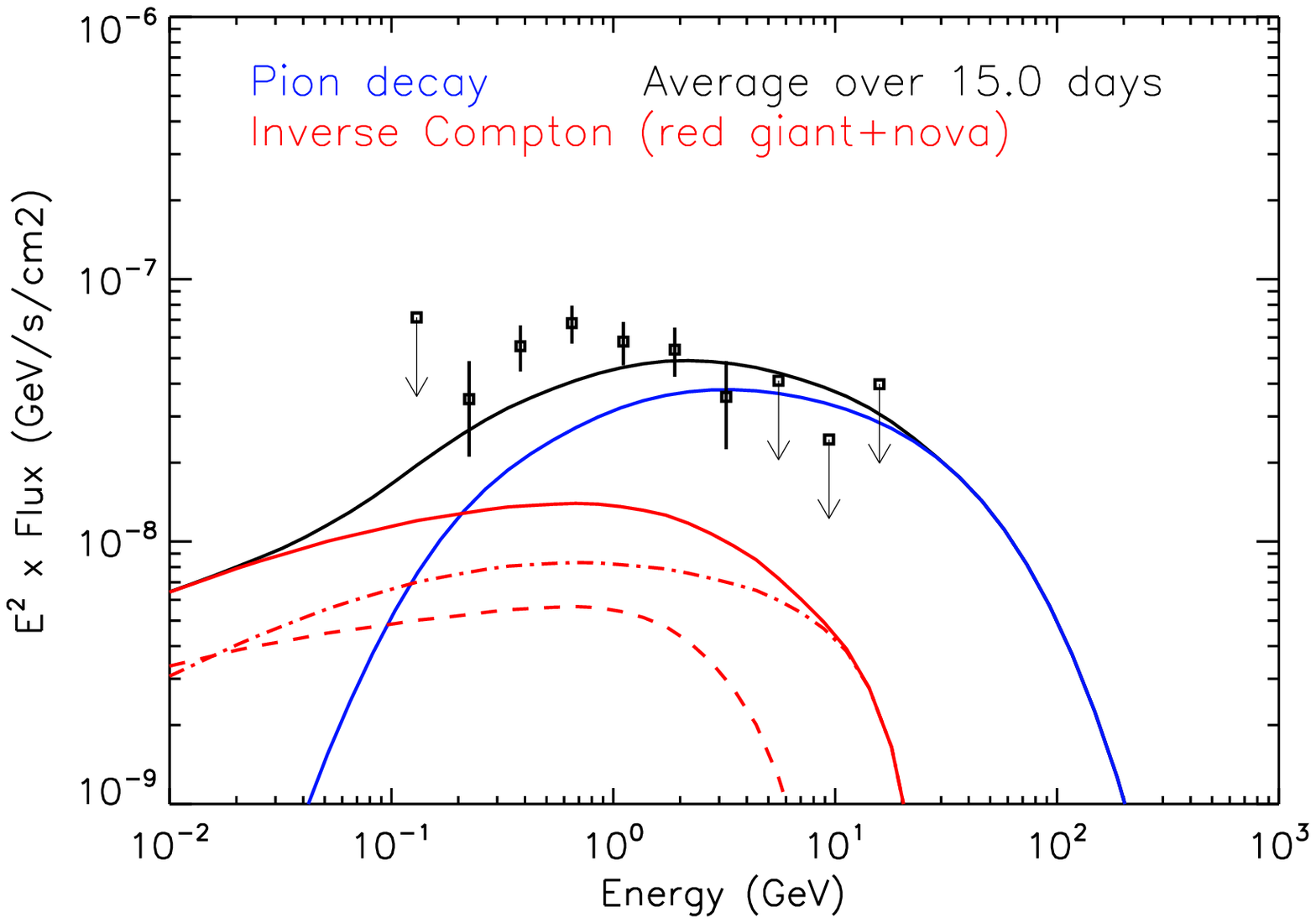}
\includegraphics[width= \columnwidth]{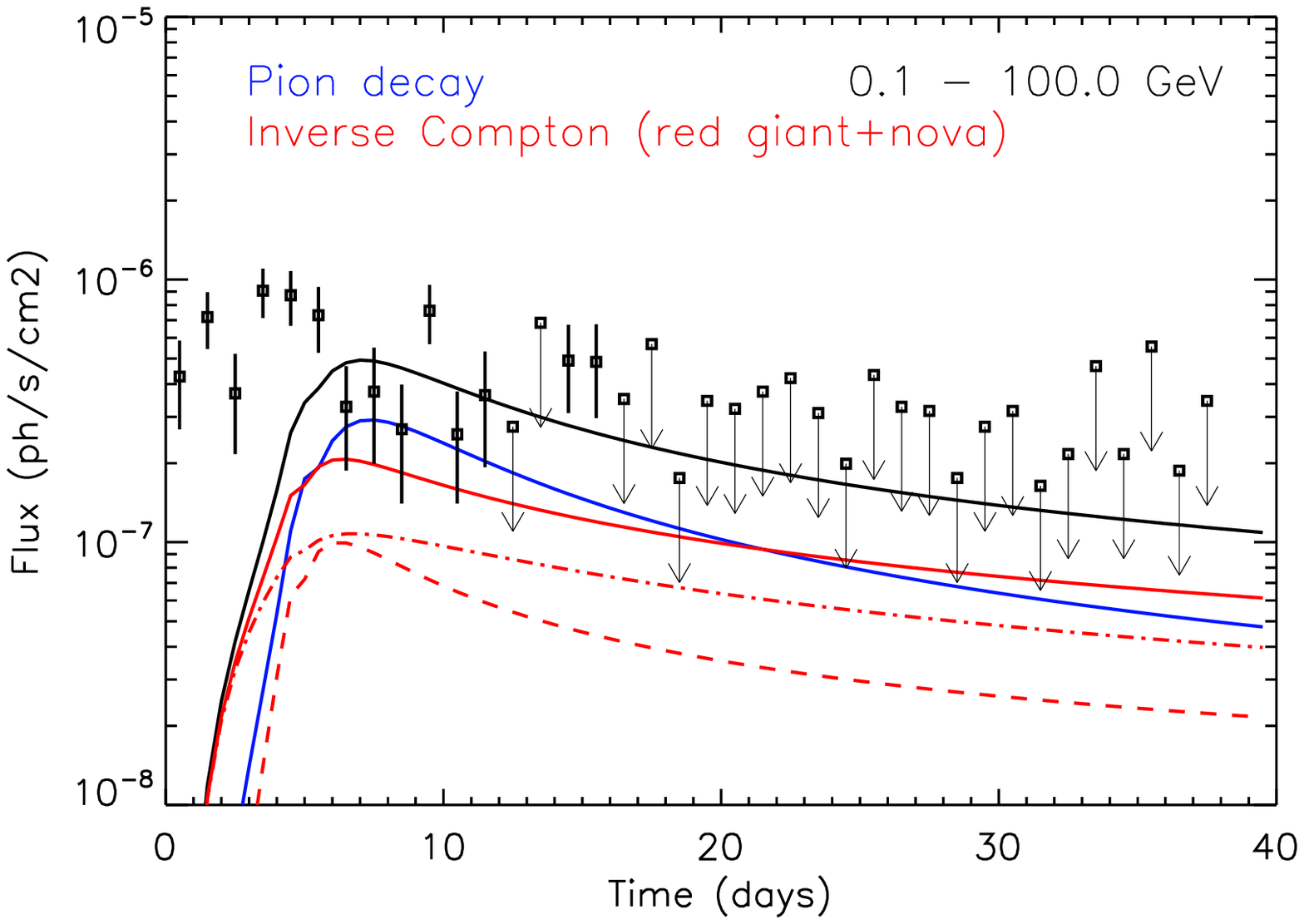}
\includegraphics[width= \columnwidth]{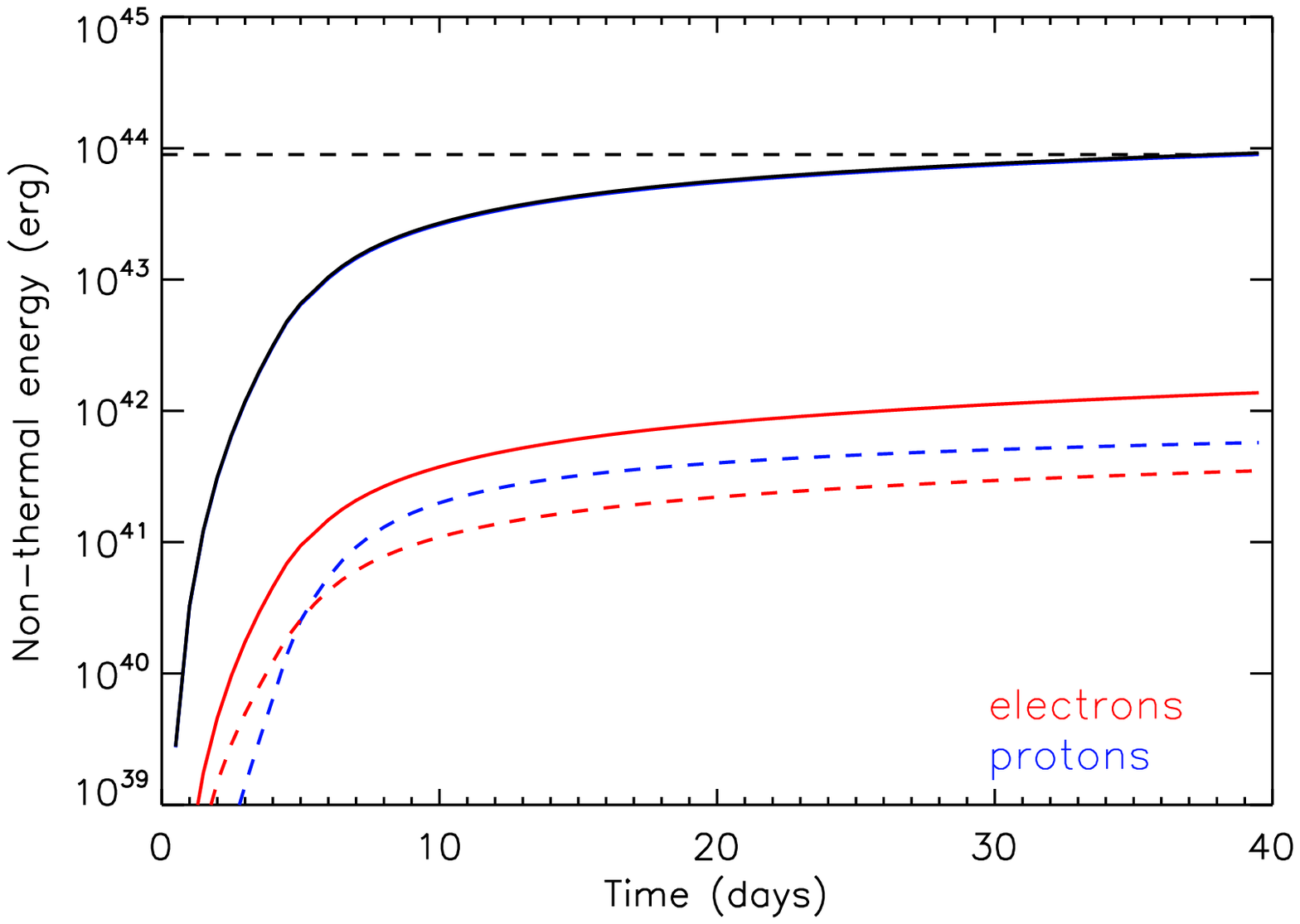}
\caption{Properties of the $\gamma$-ray emission for Run 1, the base case scenario of a shock propagating in a wind. \textit{Top panel:} time-averaged spectrum over the first 15\,days; the dashed (dot-dashed) curve is the inverse-Compton scattering on the red giant (nova) photons. \textit{Middle panel:} light curves in the 100\,MeV-100\,GeV band; the collective upper limit on the flux from day 19 to 33 is 0.8 $\times 10^{-7}$\funit\ (not shown for simplicity). \textit{Bottom panel:} non-thermal energy in particles and cumulated radiation; the red and blue solid (dashed) curves correspond to particles (radiation); the black solid (dashed) curve is the total non-thermal energy (initial nova kinetic energy).\vspace{0.5cm}}
\label{fig_emission_basecase}
\end{center}
\end{figure}

\indent Based on the non-thermal particle distributions presented in Sect. \ref{pop}, we computed the $\gamma$-ray emission of the \object{V407 Cyg} outburst due to inverse-Compton scattering of the electrons and hadronic interactions of the protons. The results are presented in Fig. \ref{fig_emission_basecase} for both populations simultaneously. The injection fractions were set at $\eta_{\mathrm{inj,p}}=3 \times 10^{-3}$ and $\eta_{\mathrm{inj,e}}=3 \times 10^{-5}$, in order to match the level of the $\gamma$-ray emission. We discuss below the properties of these emissions and compare them to the {\em Fermi}/LAT data. We emphasise that this discussion is really specific to the injection fractions and binary system parameters adopted in the base case scenario.\\
\indent \textit{Spectrum}: Below 150\,MeV, the emission is contributed in similar proportions by inverse-Compton on the nova light and RG light. Above 150\,MeV, the emission associated with neutral pion decay dominates over that from inverse-Compton. The total time-averaged spectrum exhibits a broad maximum at about a few GeV and exceeds the highest-energy points from the {\em Fermi}/LAT measurement. This results from a non-thermal particle population extending too high in energy. A lower momentum cutoff in the particle distribution would sharpen and shift the spectral profile to lower energies, making it more consistent with the observations. This could be obtained with an enhanced diffusion efficiency, as shown below.\\
\indent \textit{Light curve}: The time-averaged photon flux is at about the expected level, but the predicted light curve clearly differs from the observed one. The early rise is contributed almost exclusively by inverse-Compton on the nova light, but at too slow a rate. The flux maximum is reached at about day 6-7, while $\gamma$-ray observations indicate a maximum at day 3-4. Moreover, the light curve does not fall as rapidly as observed: at day 40, the predicted photon flux has dropped by a factor of $\simeq$5 from its peak value, whereas $\gamma$-ray observations showed a decrease by more than an order of magnitude (the collective upper limit on the emission from day 19 to 33 is 0.8 $\times 10^{-7}$\funit, not shown to not overload the plots). The emission peak occurs as the shock moves through the regions of highest matter and radiation density, which in the base case scenario are found close to the RG surface. The subsequent shallow decrease occurs because as the shock moves away from the binary, the steady increase of the number of particles with energies of $\sim$10-100\,GeV (those responsible for $\gamma$-ray emission in the $\sim$1-10\,GeV range) partially compensates the decreasing matter and radiation density.\\
\indent \textit{Locus}: Figures \ref{fig_emax_map} and \ref{fig_ntnrj_map} show that the maximum momentum and energy budget of accelerated particles vary modestly over a sizable fraction of the shock front. In contrast, the high-energy emission is contributed by a limited part of the blast wave. This is illustrated by Fig. \ref{fig_fgamma_map} for the inverse-Compton and pion decay radiation processes. Both emissions are concentrated in the vicinity of the star, where the number of accelerated particles and the density of target gas and photons are the largest. Due to its higher radiative efficiency, however, the leptonic emission arises from a more extended domain. This explains why the pion decay light curve drops slightly faster than the inverse-Compton one. This shows once again that in the base case, the flux maximum is reached approximately when the shock reaches the RG.
Figure \ref{fig_fgamma_map} also shows that the emission distribution of both the hadronic and leptonic components is truncated, suggesting that our model may underestimate the $\gamma$-ray flux because of the dead zone behind the RG. The spatial profiles seem to indicate that most of the inverse-Compton emission is captured while the pion decay emission is more impacted by the dead zone truncation. Trying to visually extrapolate the spatial profiles into the dead zone, it seems reasonable to consider that the missing contribution is $<$50\% for dominantly hadronic scenario, and $<$10\% for dominantly leptonic scenario (such as the one that eventually is considered as the best option, see Sect. \ref{gam_run4}). Yet, providing quantitative estimates of the missing emission contribution is tricky because the dead zone does not only suppress a fraction of the shock front, it also hides potentially important features of the phenomenon such as the convergent flow behind the RG, which may result in efficient particle acceleration and radiation.\\
\indent \textit{Energetics}: With the injection fractions adopted, the total acceleration efficiency at day 40 in the base case is about 100 \%, with protons largely dominating the non-thermal energy budget. The non-thermal energy output to radiation is about the same for protons and electrons, despite the former being far more numerous than the latter. This is due to the higher radiation efficiency of high-energy electrons. The bottom plot of Fig. \ref{fig_emission_basecase} shows that protons radiate inefficiently even in the dense RG wind and retain most of their non-thermal energy (only $\simeq$1\% has been converted in radiation after 40 days), while electrons lose their non-thermal energy quite rapidly (about $\simeq$20\% have been converted in radiation after 40 days).\\
\indent Overall, with the base case hypotheses, our model gives a $\gamma$-ray emission that is at approximately the observed level, and the $\gamma$-ray energies are about the measured ones for Bohm diffusion in the wind equipartition magnetic field. The predicted emission, however, reaches its flux maximum too late and does not decay rapidly enough. More problematic are the non-thermal energetics: at the time of the predicted emission maximum the acceleration efficiency is a reasonable $\simeq$20\%, but the continuation of the acceleration process leads at day 40 to a total non-thermal energy in excess of the initial nova kinetic energy. Potential solutions to this energy problem could be a larger ejecta mass, which means a larger kinetic energy reservoir, or a higher electron-to-proton ratio at injection $K_{\mathrm{ep}}$, where the proton injection would be reduced to not put a strain on the energy budget, while the electron injection would be increased to provide enough $\gamma$-rays. Yet, a more prohibitive shortcoming is the associated thermal emission in X-rays at late times (not shown), which is about 30 times above the luminosity maximum inferred from {\em Swift}/XRT observations. This means that the medium density is too high and that too much material was swept-up by the blast wave over the first three weeks. If the environment just consists of the RG wind, the stellar mass-loss rate should be $\dot{M}_{\mathrm{RG}} \leq 10^{-7}$\wunit\ in order to get thermal emission at the right level. In the next subsection, we investigate how changing the main parameters of the system impact the emission properties, and we search for a combination of plausible values that could lead to a better agreement with the observations.

% Shock propagating in a wind: parameter dependence
\subsection{Shock propagating in a wind: parameter dependence}
\label{gam_run2}

\indent The initial shock speed can be considered to be established with relatively high accuracy from emission line observations, but most other parameters have uncertainties reaching about an order of magnitude in some cases. Below, we examine their effects on particle acceleration and non-thermal emission, changing only one parameter at a time from the base case scenario.\\
\indent \textit{Orbital separation}: Decreasing the orbital separation means a shock starting its propagation in higher gas densities. The ED-stage is shorter and the shock velocity drops faster in ST-stage, but in terms of particle acceleration this is compensated by stronger magnetic fields, hence smaller diffusion coefficient and higher acceleration rate. Increasing the orbital separation leads to exactly the opposite. We tested three values $d_{\mathrm{orb}}$=5,10,15\,AU, and in all cases about the same maximum energy is reached by non-thermal particles, and about the same number of particles have been accelerated at day 40. As to the $\gamma$-ray emission, the main difference is in the light curve: for the smaller orbital separation case, the maximum particle energy is reached sooner and closer from the star; as a consequence, the light curve reaches its maximum sooner and the peak is more pronounced. A larger orbital separation leads to the opposite. Apart from that, all three values of $d_{\mathrm{orb}}$ yield about the same average spectrum, and a relatively flat light curve past maximum.\\
\indent \textit{Mass-loss rate}: The density encountered by the shock depends linearly on the mass-loss rate. We tested several values around $10^{-7}$\wunit. A lower mass-loss rate means a longer ED-stage but a lower magnetic field, and this eventually gives a similar maximum particle energy. Yet, with a lower mass-loss rate, higher-energy particles are produced at a greater distance from the binary system, and are taken further out at a greater velocity. This means that they are exposed to lower radiation and gas densities and for a shorter time, which implies a lower radiation efficiency. The low density affects in particular protons, whose relative contribution to the $\gamma$-ray emission can fall below that of electrons. In addition, a lower wind density for a same injection fraction also results in fewer particles being accelerated and a lower acceleration efficiency at a given time. In such conditions, reproducing a given non-thermal emission level at early times requires an increase of the injection fractions, and may lead to unrealistically high acceleration efficiencies at late times. In terms of $\gamma$-ray emission, a lower mass-loss rate leads to a sharper and earlier light curve maximum, with a more rapid decline but a smaller flux overall (for a same injection fraction). A higher mass-loss rate leads to the opposite\footnote{There is a certain degeneracy between the mass-loss rate and the wind velocity in the sense that the same density profile can be obtained through different combinations of both parameters.}.\\
\indent \textit{Ejecta mass}: The ejecta mass linearly sets the kinetic energy reservoir from which non-thermal components are powered. Too small a mass requires too high an acceleration efficiency to account for a given level of $\gamma$-ray emission, and we therefore considered only ejecta masses greater than 10$^{-6}$\msol. A higher ejecta mass implies higher shock velocities on average, because the ST-stage slowing-down occurs later and on a longer time scale. At a given time and for the same injection fractions, more particles have been accelerated and they reach higher energies because acceleration at full shock speed proceeded on a longer time. In addition, accelerated particles are advected away more rapidly, towards smaller gas and radiation densities. Despite these effects, the global acceleration efficiency is lower. The $\gamma$-ray emission in the larger ejecta mass case features a more energetic spectrum, a higher luminosity, and a more pronounced light curve maximum. This light curve maximum is, however, reached at about the same time, and the subsequent emission decrease remains nearly as flat.\\
\begin{figure}[t]
\begin{center}
\includegraphics[width= \columnwidth]{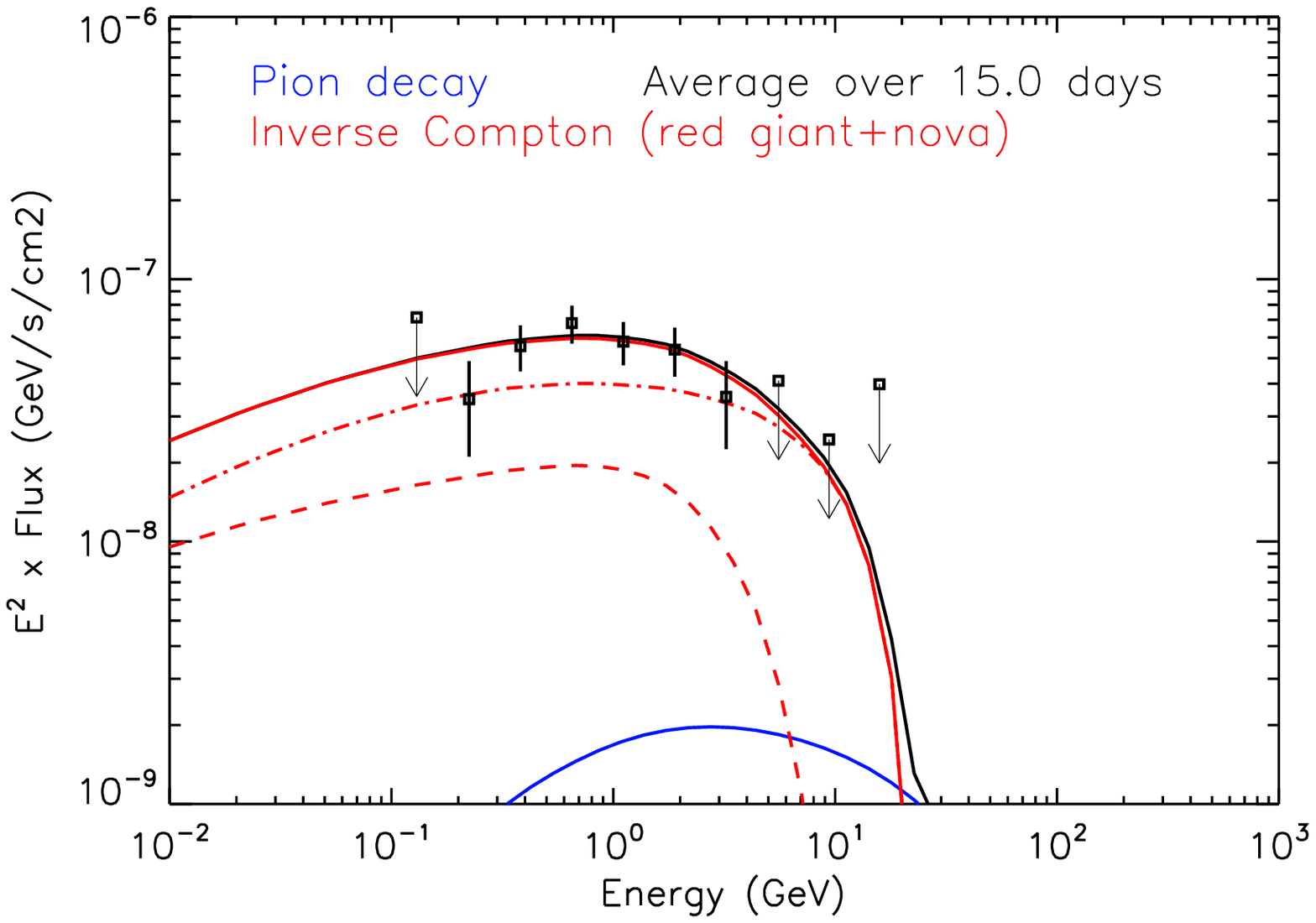}
\includegraphics[width= \columnwidth]{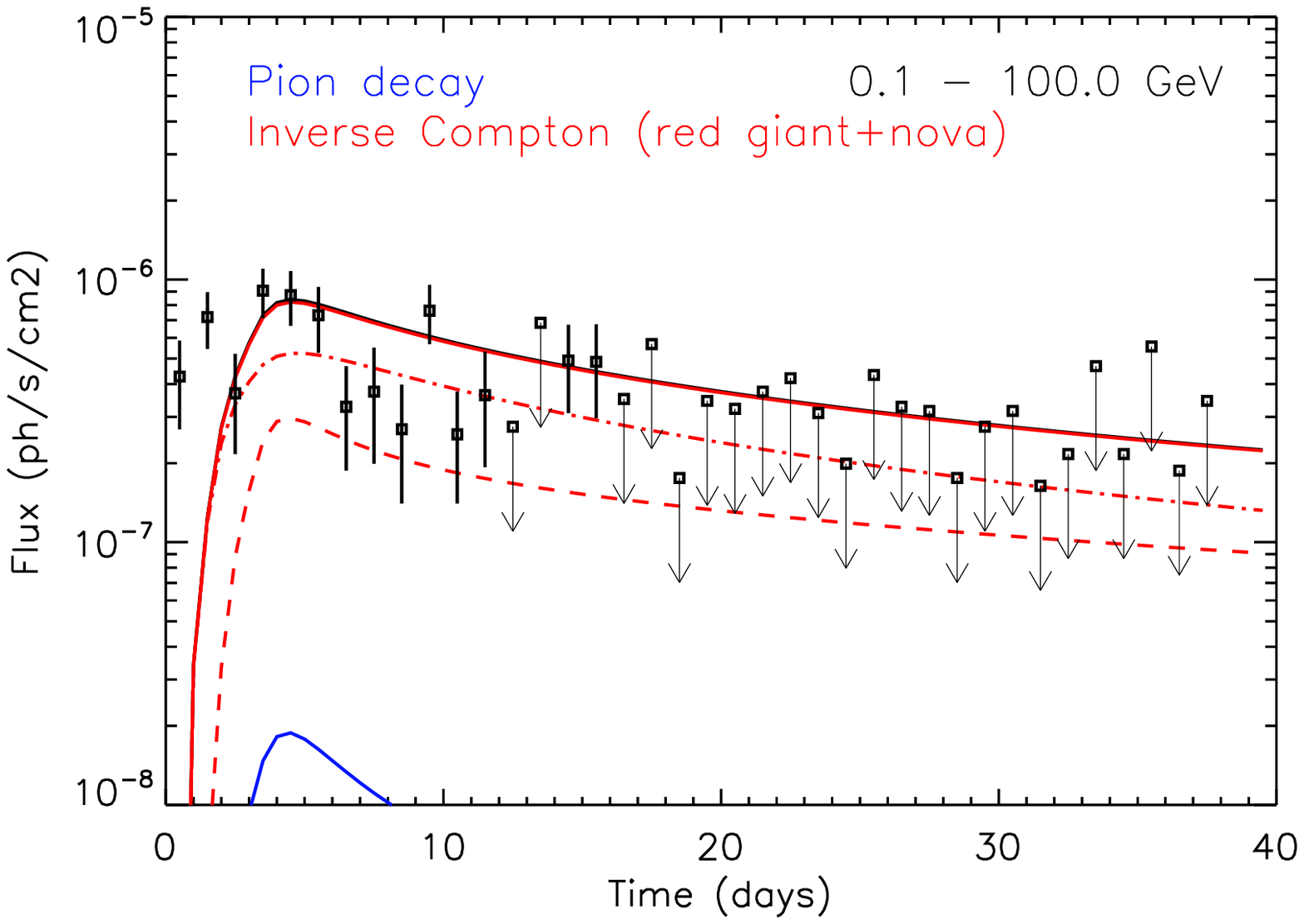}
\includegraphics[width= \columnwidth]{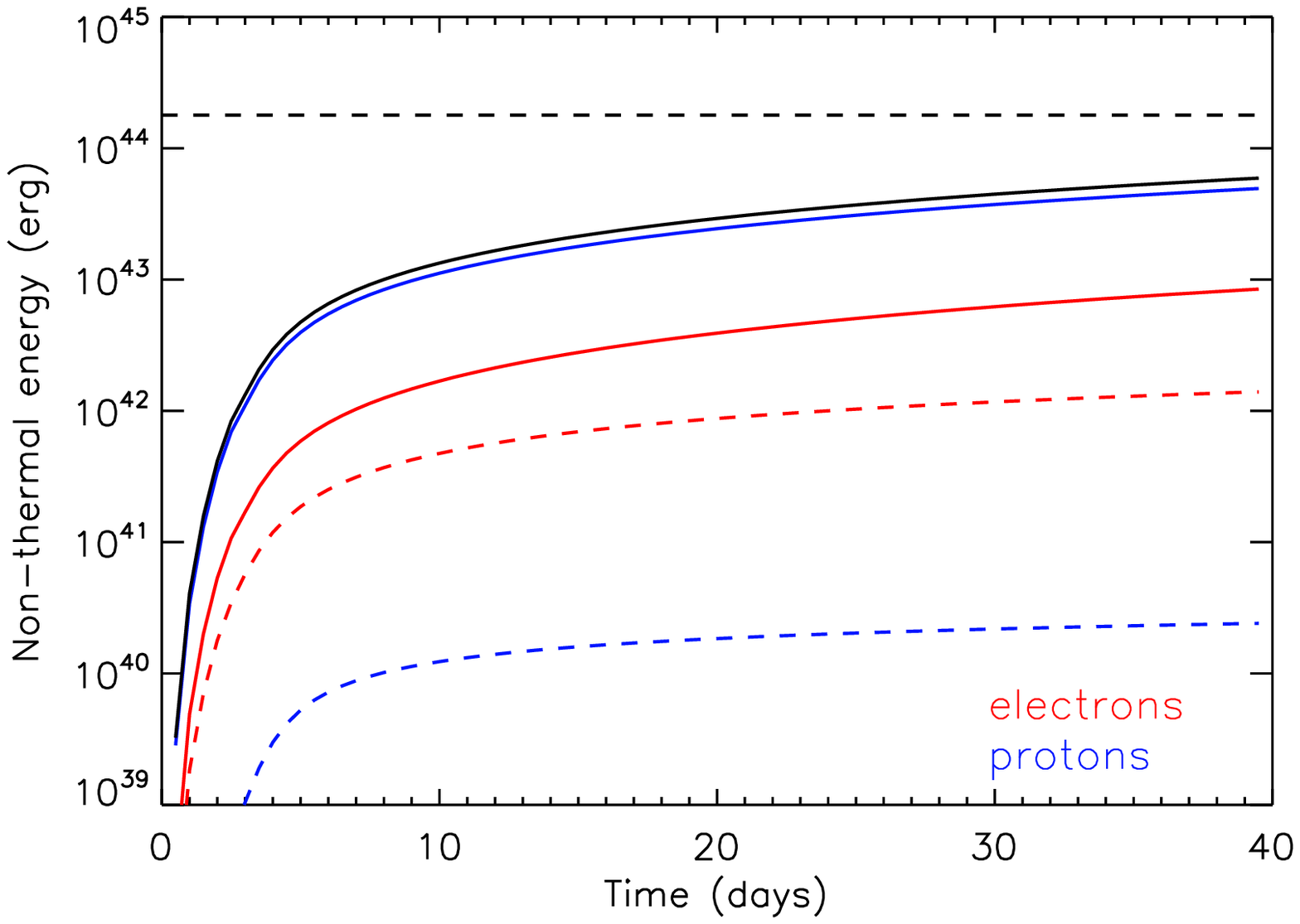}
\caption{Properties of the $\gamma$-ray emission for Run 2, the scenario of a shock propagating in a wind, optimized for the case of V407 Cyg. Color coding and graph elements are described in Fig. \ref{fig_emission_basecase}.}
\label{fig_emission_optimised}
\end{center}
\end{figure}
\indent \textit{Diffusion coefficient}: The base case calculation was based on the assumption of Bohm diffusion (in a non-amplified upstream field), which provides the maximum acceleration rate. This prescription can be relaxed through the parameter $\zeta$ in Eq. \ref{eq_diff}, which allows scaling up the scattering mean free path as a multiple of the Larmor radius. When energy losses can be neglected (as for protons), this results in a maximum particle energy that is lower by a factor $\zeta$; but if energy losses play a major role (as for electrons), there is still a reduction but not by $\zeta$ because the maximum energy is then set by a balance between acceleration and energy losses. In terms of energetics, increasing the diffusion coefficient decreases the total non-thermal particle energy. Proportionally, this mostly affects protons because they are age-limited while electrons are loss-limited. The consequences of a moderately higher diffusion coefficient on the $\gamma$-ray emission are mainly a spectrum shifted to lower energies, while the light curve is left almost unchanged\footnote{The effect of $\zeta$ can also be mimicked through a modification of the wind temperature, since this partially defines the equipartition magnetic field (the implications on the synchrotron emission are different though).}.

% Shock propagating in a wind: optimised scenario
\subsection{Shock propagating in a wind: optimised scenario}
\label{gam_run3}

\indent From the above discussion, the trend to follow in order to get a better agreement with observations would be (compared to the base case scenario): higher ejecta mass, smaller orbital separation, larger diffusion coefficient, and smaller mass-loss rate. These choices favour a rapid rise and decline of the light curve, a global acceleration efficiency down to moderate values, a lower $\gamma$-ray spectral maximum, and thermal emission at the level observed in X-rays. We tested several possibilities along this trend and eventually identified an optimised scenario that we will refer to as Run 2, and whose parameters are listed in Table \ref{tab_run}.\\
\indent The lower wind densities imposed by thermal emission constraints have two direct implications: a reduced proton radiation efficiency, and fewer particles being accelerated for a given injection fraction. To obtain $\gamma$-ray emission at a given level, these effects can be compensated by a higher proton injection fraction, but that means an enhanced consumption of the nova kinetic energy. In Run 2, getting pion decay emission at the observed level requires $\eta_{\mathrm{inj,p}}=3 \times 10^{-1}$ and a non-thermal efficiency above 100\% after just a few days. Overall, with such density conditions, the $\gamma$-ray emission cannot be of hadronic origin. A leptonic origin can be considered, but only at the expense of a higher $K_{\mathrm{ep}}$: agreement with the data can be obtained for $\eta_{\mathrm{inj,p}}=6 \times 10^{-3}$, $\eta_{\mathrm{inj,e}}=6 \times 10^{-4}$, as illustrated in Fig. \ref{fig_emission_optimised}.\\
\indent The predicted time-averaged spectrum is fully consistent with the measured one, and this is obtained for particle scattering at the Bohm limit, with $\zeta=1$. The predicted light curve reaches its maximum at day 3-4, as observed, but falls short of the emission detected over the first two days. Actually, the rising part of the light curve is consistent with the data if one takes into account the 3-day uncertainty in the exact date of the outburst. The predicted emission, however, does not decline fast enough beyond the first two weeks and exceeds the collective upper limit at late times by a factor of 2. All this is achieved with a global acceleration efficiency of $\simeq30\%$ at day 40, but still on the rise so that it reaches $\simeq50\%$ at day 80.\\
\indent The above scenario yields about the best possible fit to the $\gamma$-ray observations for a shock propagating in a wind density profile. The level of thermal X-ray emission at late times, after three weeks, matches that observed with {\em Swift}/XRT. The adopted ejected mass is in the range of acceptable values for a massive WD with $M_{\mathrm{WD}} \geq 1.25$ \citep{Yaron:2005}. The adopted orbital separation is smaller than the commonly-accepted values: about 16\,AU in \citet{Munari:1990} and $\geq$12\,AU in \citet{Shore:2011}. Yet, a smaller separation cannot be completely excluded considering the uncertainties on the orbital period (Ulisse Munari, private communication). Interestingly, some features of the predicted $\gamma$-ray emission proved to be rather robust to order-of-magnitude changes of the model parameters. First, obtaining a spectrum in agreement with the observed one generally required a particle scattering close to the Bohm limit ($1 \leq \zeta \leq 10$). As will be discussed below, this is a useful element for the assessment of novae in symbiotic systems as a population of high-energy $\gamma$-ray emitters. For \object{V407 Cyg}, another robust feature of all models that have been tested is that the predicted emission never exhibited the strong drop observed by {\em Fermi}/LAT about two weeks after the initial detection. This calls for the replacement or sophistication of some of the assumptions adopted so far. In the following part, we show that taking into account a more realistic ambient medium improves the fit to the observations.

% Shock propagating in a wind with density enhancement
\subsection{Shock propagating in a wind with density enhancement}
\label{gam_run4}

\begin{figure}[t]
\begin{center}
\includegraphics[width= \columnwidth]{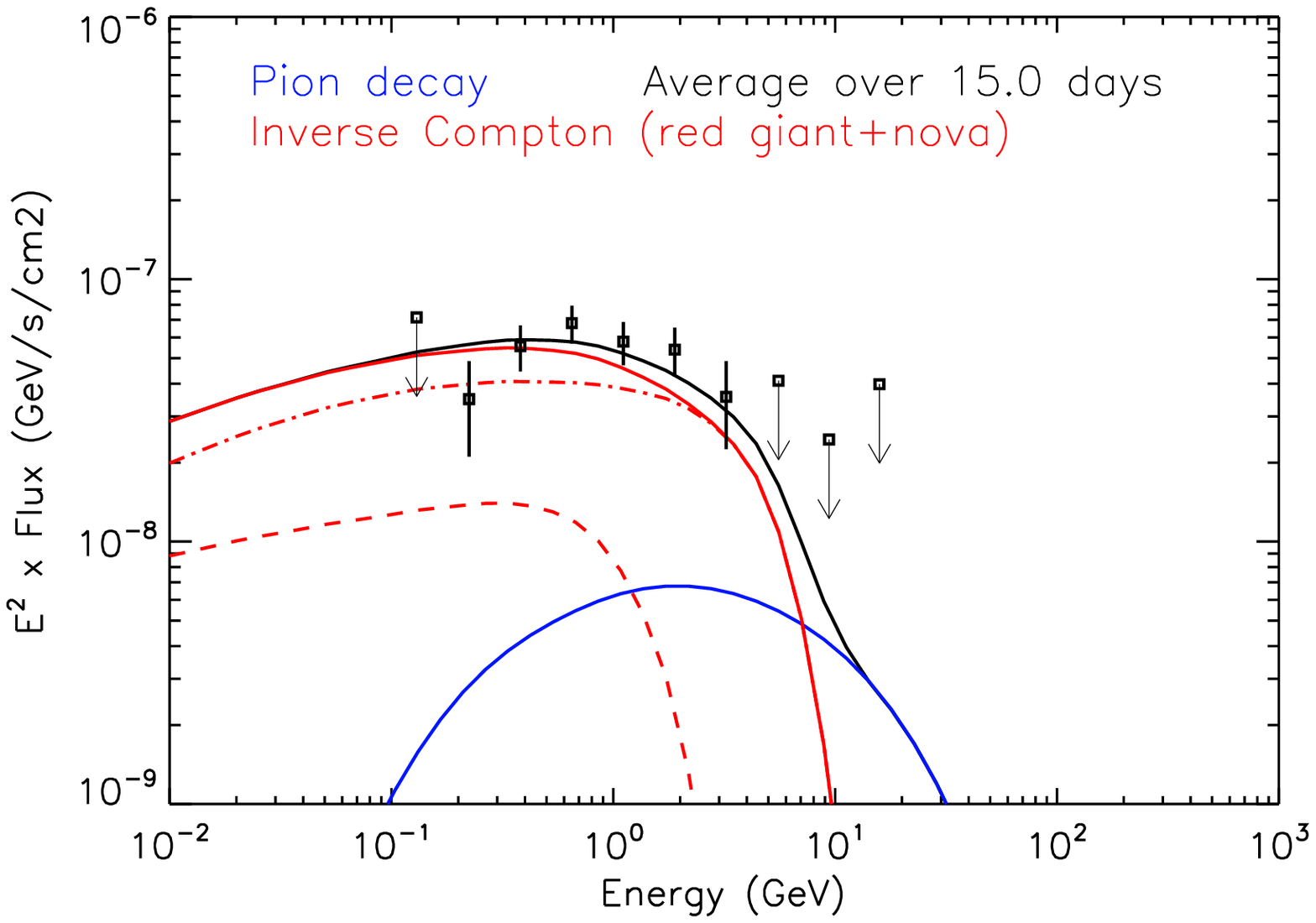}
\includegraphics[width= \columnwidth]{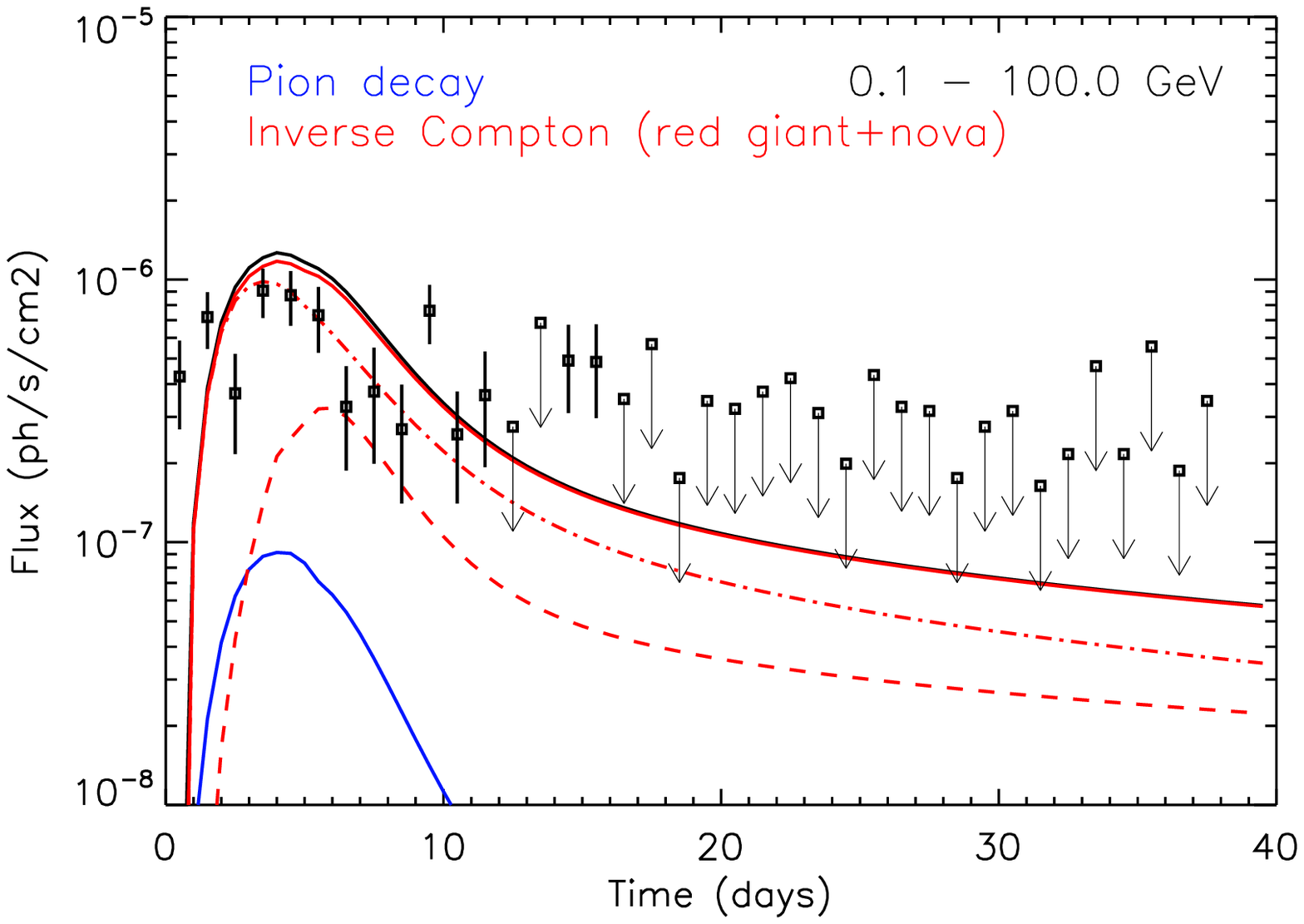}
\includegraphics[width= \columnwidth]{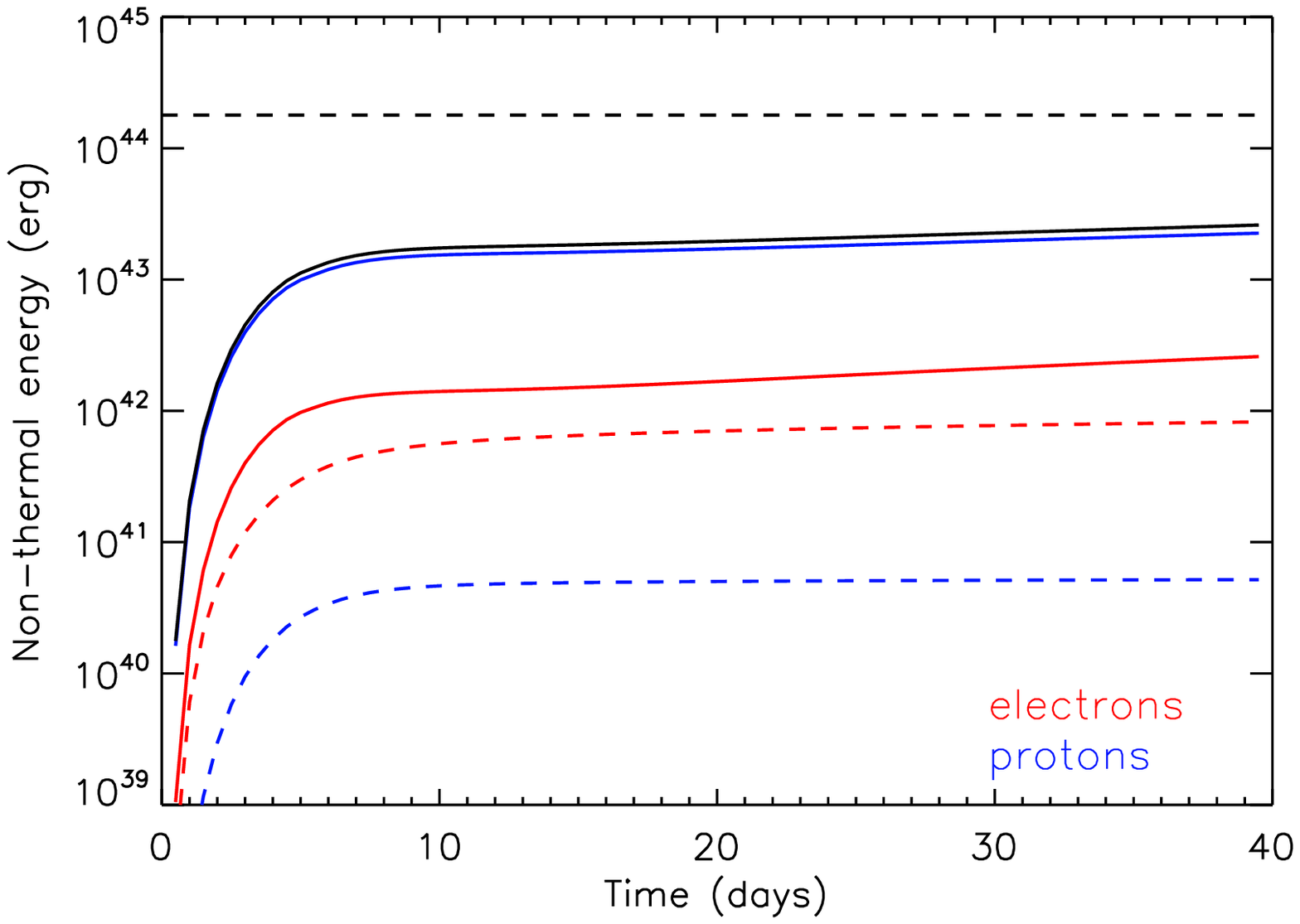}
\caption{Properties of the $\gamma$-ray emission for Run 3, the scenario of a shock propagating in a wind and density enhancement around the WD, optimized for the case of V407 Cyg. Color coding and graph elements are described in Fig. \ref{fig_emission_basecase}.}
\label{fig_emission_cde}
\end{center}
\end{figure}
\begin{figure}[t]
\begin{center}
\includegraphics[width= 8cm]{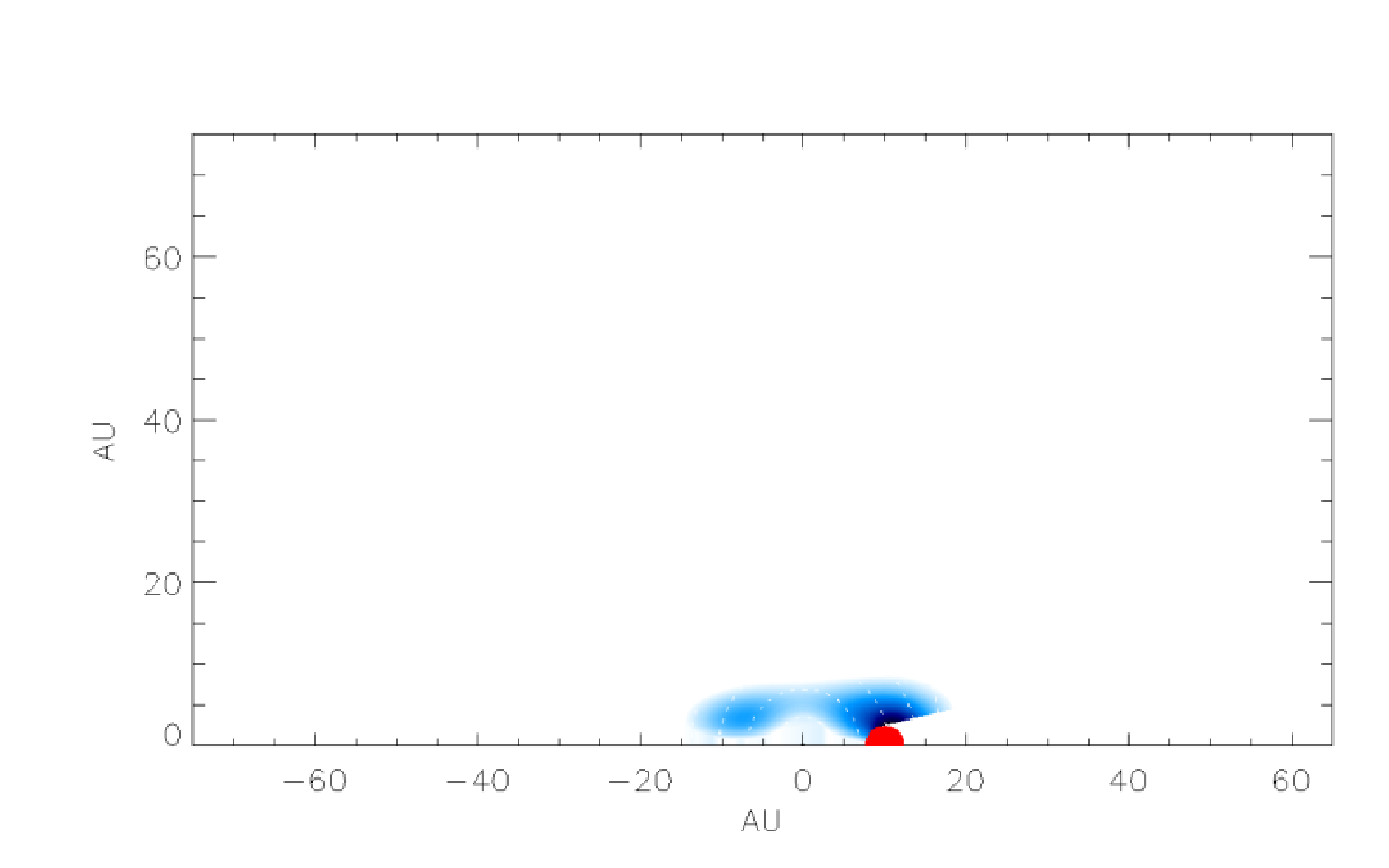} 
\includegraphics[width= 8cm]{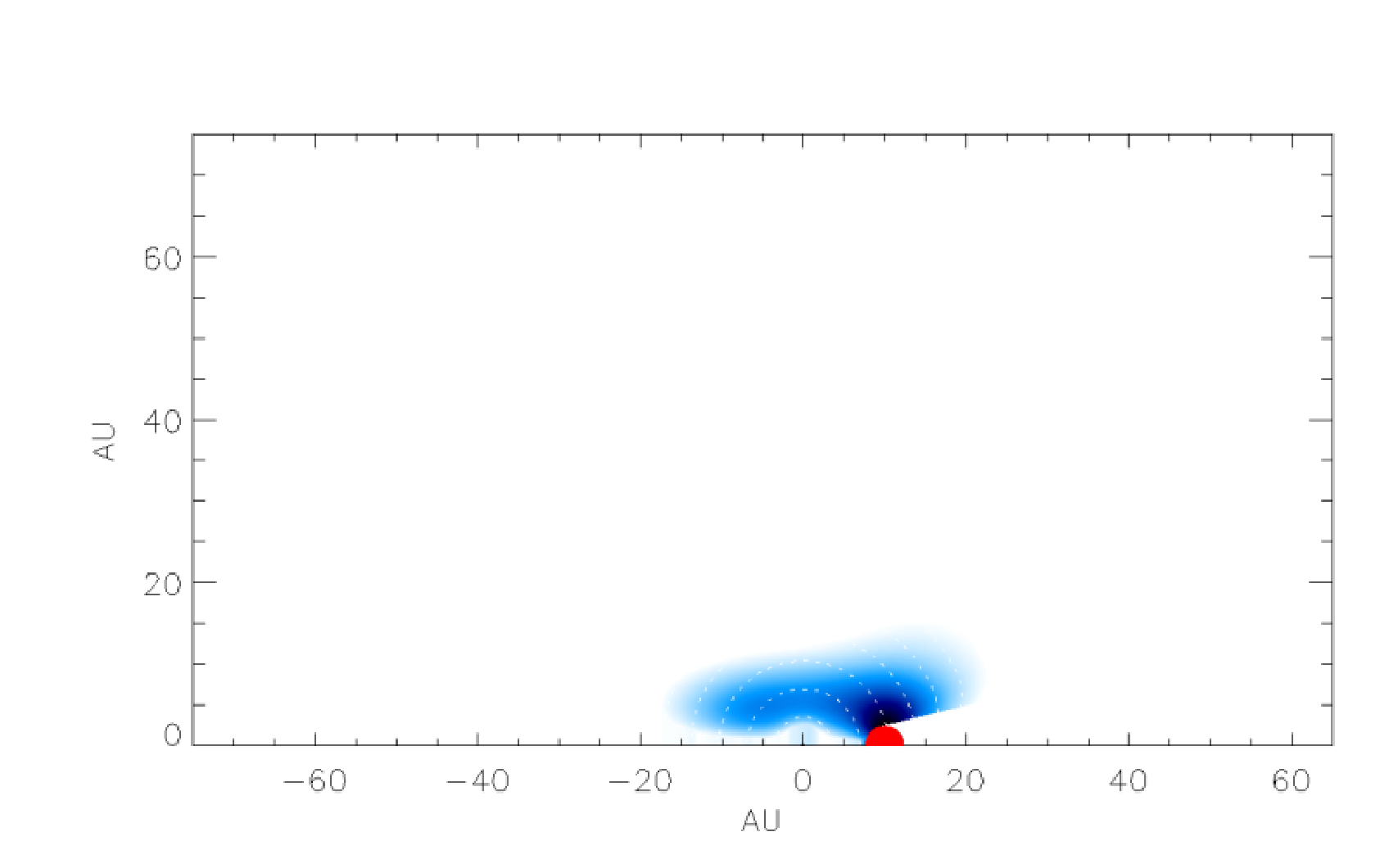}
\caption{Spatial maps of the $>$100\,MeV $\gamma$-ray emission from pion decay (top) and inverse-Compton (bottom) for Run 3, the scenario of a shock propagating in a wind and density enhancement around the WD. Color coding and graph elements are described in Fig. \ref{fig_emax_map}.}
\label{fig_fgamma_map2}
\end{center}
\end{figure}

\indent As shown above, a simple spherically-symmetric wind profile prevents a quick rise of the $\gamma$-ray emission, unless the orbital separation is small, and in any case results in a shallow decrease of the light curve past maximum. The latter shortcoming is connected with another problem, which is the steadily increasing non-thermal efficiency at late times. In this section, we mainly investigate the effect of alternative, more realistic density distributions.\\
\indent From hydrodynamic simulations of the symbiotic \object{RS Oph} system, \citet{Walder:2008} have shown that the quiescent accretion phase between outbursts leads to a structured mass distribution with enhancement in the orbital plane. This favours a poleward propagation of the shock when the system enters in outburst. This conclusion and its applicability to the \object{V407 Cyg} system were corroborated by \citet{Orlando:2012}, who found that a density enhancement on top of the wind density profile is needed to account for the X-ray light curve (see Sect. \ref{xraysynch}). Due to the cylindrically-symmetric nature of our model, we could not explore the shock propagation and associated non-thermal emission in a realistic three-dimensional matter distribution. The density enhancement was therefore modelled according to Eq. \ref{eq_CDE}, with the centre on the WD in agreement with the simulations of \citet{Walder:2008} that show a decreasing density in any direction from the WD (see their Fig. 3). This results in a cigar-like rather than a disc-like structure. Such a density profile does not allow reproducing a bilobate propagation of the shock and instead leads to a preferentially azimuthal expansion of the blast wave. This mimics the effect of more material being swept up by a fraction of a shock front and for the thermal X-ray emission at least, this turned out to be a very good approximation \citep{Orlando:2012}.\\
\indent The effect of such a circumstellar component can easily be anticipated. It will provide more material for acceleration and interaction over just a fraction of the shock front, in particular between the WD and the RG. This will allow for an earlier rise of the $\gamma$-ray emission, and for a more pronounced drop when the shock exits the density enhancement. The mass accumulated around the WD is constrained by thermal emission observations, since this additional material swept-up by the shock will shine in X-rays. We tested several possibilities and found that a CDE with a typical size of 10\,AU and a peak density of $10^{8}$\,cm$^{-3}$ provides a significantly better fit to the $\gamma$-ray data. This new configuration will be referred to as Run 3, and the corresponding parameter set is given in Table \ref{tab_run}. Compared to the previous Run 2, the density enhancement allowed to come back to a larger orbital separation and to further decrease the wind density through a reduced mass-loss rate.\\
\indent Even with a matter structure around the WD, however, reproducing the observed $\gamma$-ray emission with hadronic interactions requires high proton injection fractions and still poses an energetic problem, with acceleration efficiencies being close to 100\% for some parts of the shock (those propagating towards the RG). So once again only a leptonic option can be considered, still at the expense of a high $K_{\mathrm{ep}}$. Figures \ref{fig_emission_cde} and \ref{fig_fgamma_map2} show the $\gamma$-ray emission properties for Run 3 using $\eta_{\mathrm{inj,p}}=5 \times 10^{-3}$, $\eta_{\mathrm{inj,e}}=3 \times 10^{-4}$.\\
\indent The predicted $\gamma$-ray spectrum required again particle scattering close to the Bohm limit with $\zeta=3$ in order to match the measured one. The emission is mostly of leptonic origin and consists primarily of inverse-Compton scattering on the nova light. Hadronic emission dominates only above 10\,GeV but in the downturn part of the pion bump, so significant emission in the 100\,GeV range and beyond is not expected. In terms of spectral variability over the first 15 days, the predicted spectrum initially has a cutoff at $\sim$100\,MeV and extends up to $\sim$1\,GeV in 3 days. The spectral shape is then relatively stable over the following 10 days. At later times, 30-40 days, the spectrum is harder because the electron distribution flattens at high energy, as the inverse-Compton loss rate drops over time (see Fig. \ref{fig_distrib_basecase_angle}). Overall, our result are consistent with the absence of spectral variability observed by \citet{Abdo:2010c} for two equivalent statistical significance time intervals.\\
\indent The predicted light curve reaches its maximum at day 3-4, but the rise is not fast enough to account for the emission measured over the first day of {\em Fermi}/LAT observations. The peak of the $\gamma$-ray light curve actually is due to the combined action of the density enhancement and nova, the former providing sufficient electrons for acceleration and inverse-Compton emission on the photons of the latter, when the shock is still near the WD. As soon as the shock leaves the density enhancement and the vicinity of the binary, the emission drops and acceleration proceeds in the lower-density wind. The flux decreases by a factor 20 in the few weeks that follow the emission maximum, as observed. The global non-thermal efficiency is about $\simeq$10\% after two weeks, and subsequently increases at a modest rate of about 1\% per week (at least up to day 40). The predicted light curve slightly overshoots the highest {\em Fermi}/LAT data points, and part of this deviation arises from the different nature of what we are comparing. The {\em Fermi}/LAT data points were actually determined using the exponentially-cutoff power-law fit to the time-averaged data, while our model gives the total 100\,MeV-100\,GeV photon flux without any assumption on the spectral shape. If we derive the light curve by fitting the {\em Fermi}/LAT spectrum to our model spectra (in the 100\,MeV-5\,GeV range), we obtain $\sim$20\% lower photon fluxes because our model is flatter at low energies $\sim$100\,MeV than what is allowed by the exponentially-cutoff power-law shape. Also, some data points are still at 2-3$\sigma$ off the predicted light curve, notably at days 1, 6, 16, and 17. With a possible 1-day offset between the outburst of the nova and its first optical detection, the early part of the predicted light curve fits rather satisfactorily the data; then, in addition to statistical fluctuations, the deviations may be due to inhomogeneities in the density profile of the surrounding medium and/or clumpiness of the ejecta \citep{Williams:1994}.\\
\indent The predicted thermal X-ray luminosity after three weeks is within a factor of a few of the value inferred from {\em Swift}/XRT observations. When interpreted with the prior knowledge of the limitations of our thermal model, this result can actually be considered as satisfying (see Sect. \ref{xraysynch} and Fig. \ref{fig_xray}). We emphasise that the parameters of Run 3 were not obtained through a formal multivariate fit under the constraints from X-ray and $\gamma$-ray observations. Our aim was instead to illustrate some trends in the framework of our model, and provide typical values for the main parameters of the system. On that point, the positive influence of some sort of circumstellar density enhancement is noteworthy, but the statistical significance of the improvement brought by adding the corresponding 3 degrees of freedom to the model could not be derived. The determination of the exact shape of the CDE is beyond the capabilities of our model, and could more likely be done from the analysis of the thermal continuum and line emission through hydrodynamical simulations, in the same vein as \citet{Orlando:2012}. Within our model, the experimental data can be accommodated with a moderately more extended CDE having a lower density, or the opposite. The important quantity here is the swept-up mass, which drives the thermal X-ray output and sets the amount of material available for acceleration. Similarly, the orbital separation used is still a bit smaller than the available estimates, but the model can accommodate slightly larger values when other parameters are adjusted.\\
\indent A potentially important effect in $\gamma$-ray-emitting binaries is the anisotropic inverse-Compton scattering, and we investigated if that could improve the fit of our model predictions. Because the distribution of seed photons for the inverse-Compton scattering is not isotropic, the line-of-sight orientation with respect to the binary affects the observed $\gamma$-ray emission. We considered the extreme cases of superior conjunction (favouring head-on collisions with photons from the RG, at least while some parts of the shock are still inside the binary) and inferior conjunction (favouring trailing collisions) for the system viewed edge-on. The results for the inverse-Compton radiation are shown in Fig. \ref{fig_emission_anisoIC} for the two inverse-Compton components (the total spectra and light curves are not shown to not overload the plots). The impact of anisotropy on the nova component is modest over the first 10 days because the electron population is distributed nearly spherically about the source of seed photons. At later times, the shock geometry departs from spherical symmetry and anisotropic effects become apparent in the light curve. In contrast, there is a significant impact of anisotropy on the RG component. The case of superior conjunction yields a higher flux, but the most interesting effect is on the light curve, with a flux maximum being sharper and a subsequent decline being stronger. Interestingly, based on an analysis of several emission line profiles, \citet{Shore:2011} suggest that we are seeing the system close to superior conjunction. Yet, since the RG component is not the dominant one, the effect of anisotropic inverse-Compton on the total emission remain limited.

% X-ray and radio emission
\section{X-ray and radio emission}
\label{xraysynch}

\indent It was shown above that the $\gamma$-ray emission should be interpreted together with the thermal emission constraints, since the latter dictate the geometry of the binary system and the hydrodynamics of the shock expansion. Yet, our simple model for the thermal X-ray emission suffers from serious limitations, which restricted the amount of information we could derive from the {\em Swift}/XRT observations but still allowed to constrain the nova properties as we show below. Then, to consolidate the scenario of Run 3 presented above, and in particular the conditions of particle acceleration, we discuss its synchrotron emission in light of radio observations performed in the $\sim$1-10\,GHz range (see Sect. \ref{obs}).

% Thermal emission
\subsection{Thermal emission}
\label{xraysynch_xray}

\begin{figure}[t]
\begin{center}
\includegraphics[width= 8cm]{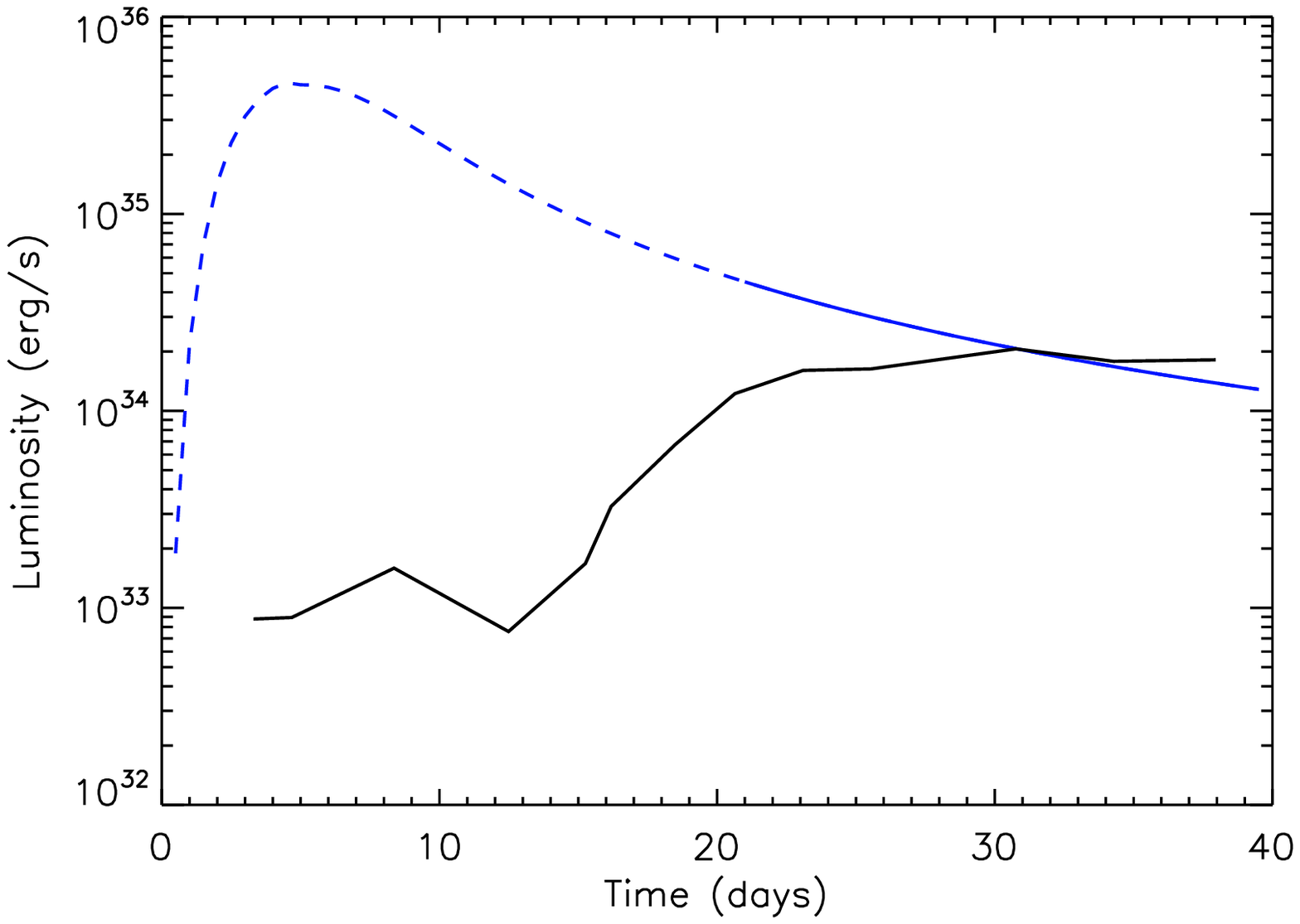} 
\caption{Predicted thermal X-ray light curve for Run 3 (in blue), compared to that inferred from {\em Swift}/XRT \citep[in black, reproduced from][]{Orlando:2012}. A comparison with the results of multi-dimensional hydrodynamical simulations showed that our model is unable to account for the complexity of the emission rise. Only the average luminosity past three weeks, the solid part of the curve, was therefore used as a constraint on the non-thermal processes.}
\label{fig_xray}
\end{center}
\end{figure}
\indent Our approach to compute the thermal Bremsstrahlung emission from shock-heated material is quite close to that adopted by \citet{Nelson:2012}, and we indeed obtained similar results to theirs when using the same description of the binary system. A more sophisticated study was done by \citet{Orlando:2012} using multi-dimensional hydrodynamical simulations including radiative cooling and thermal conduction. Comparing our results to theirs for a similar setup (for instance their best 2D run E44.3-NW7-CDE6.3-L40 including a CDE), it appears that our model for the thermal X-rays is unable to account for the emission rise over the first three weeks, as illustrated in Fig. \ref{fig_xray}. \citet{Orlando:2012} show how sensitive the emission rise is to the circumstellar/circumbinary conditions, so it is not surprising that our simple model cannot capture such details. The reasons for such a shortcoming were already pointed out in \citet[][see their Sect. 6.2]{Nelson:2012}. In addition, there are more subtle effects due to what happens on the rear side of the RG\footnote{Flow convergence on the rear side of the Mira star produces a hot and dense plasma that contributes strongly to the thermal X-ray emission. The position of the flow convergence is what determines the rise of the X-ray luminosity, and this depends on the density profile of the ambient medium; adding a density enhancement to the wind density profile introduces a delay in this convergence and this is proposed as an explanation for the late X-ray luminosity maximum (Salvatore Orlando, private communication).}, as discussed in \citet{Orlando:2012}. This is a phenomenon that our model cannot reproduce because of the dead zone behind the RG. At a later stage, however, our model gives rather good results with average X-ray luminosities that are a factor $\simeq$2 above those obtained from the full hydrodynamical simulations. At these times, the emission is likely driven more by the total amount of swept-up material than by the way in which this material was accumulated. In other words, our thermal model cannot reproduce the full development of the thermal emission. A joint fit to the X-ray and $\gamma$-ray observations is therefore not possible in the adopted framework. All the information about the density structure of the environment that is encoded in the X-ray luminosity rise is not accessible to us. Still, having a global constraint on the total amount of swept-up mass is crucial because this is the reservoir of projectiles and targets for particle acceleration and radiation. In the different scenarios discussed above, we therefore required that the average predicted X-ray luminosity past three weeks does not exceed a few times 10$^{34}$\,erg\,s$^{-1}$ in the {\em Swift}/XRT band.\\
\indent If the addition of a CDE in our model was motivated by the work of \citet{Orlando:2012}, we reached different conclusions about the properties of this gas structure. The CDE of Run 3 presented above is smaller, with a size of about 10\,AU rather than 40\,AU, and a density of $10^{8}$\,cm$^{-3}$ instead of $2 \times 10^{6}$\,cm$^{-3}$ (although our values should not be considered as precise optima, as already emphasised). We note that many of their assumptions differ from those of our Run 3, and this is probably what led to this discrepancy: they used an initial ejecta velocity of about 9000\kms, while we adopted 3000\kms; their ejecta mass is 2 $\times$ 10$^{-7}$\msol, which is an order of magnitude below our 2 $\times$ 10$^{-6}$\msol; last, they considered an orbital separation of 15.5\,AU where we used 10\,AU. With such parameters, we found it almost impossible to account for the $\gamma$-ray observations with reasonable parameters. Owing to the low densities, that could be only be done with dominantly leptonic emission. Yet, the electron acceleration efficiency would be about 10\% and the electron-to-proton at injection close to 1. The $\gamma$-ray emission maximum is reached early enough, but the subsequent light curve drop is far too slow. In the framework of our model, the scenario considered by \citet{Orlando:2012} for the thermal emission does not fit the non-thermal constraints.

% Synchrotron emission
\subsection{Synchrotron emission}
\label{xraysynch_synch}

\begin{figure}[t]
\begin{center}
\includegraphics[width= 8cm]{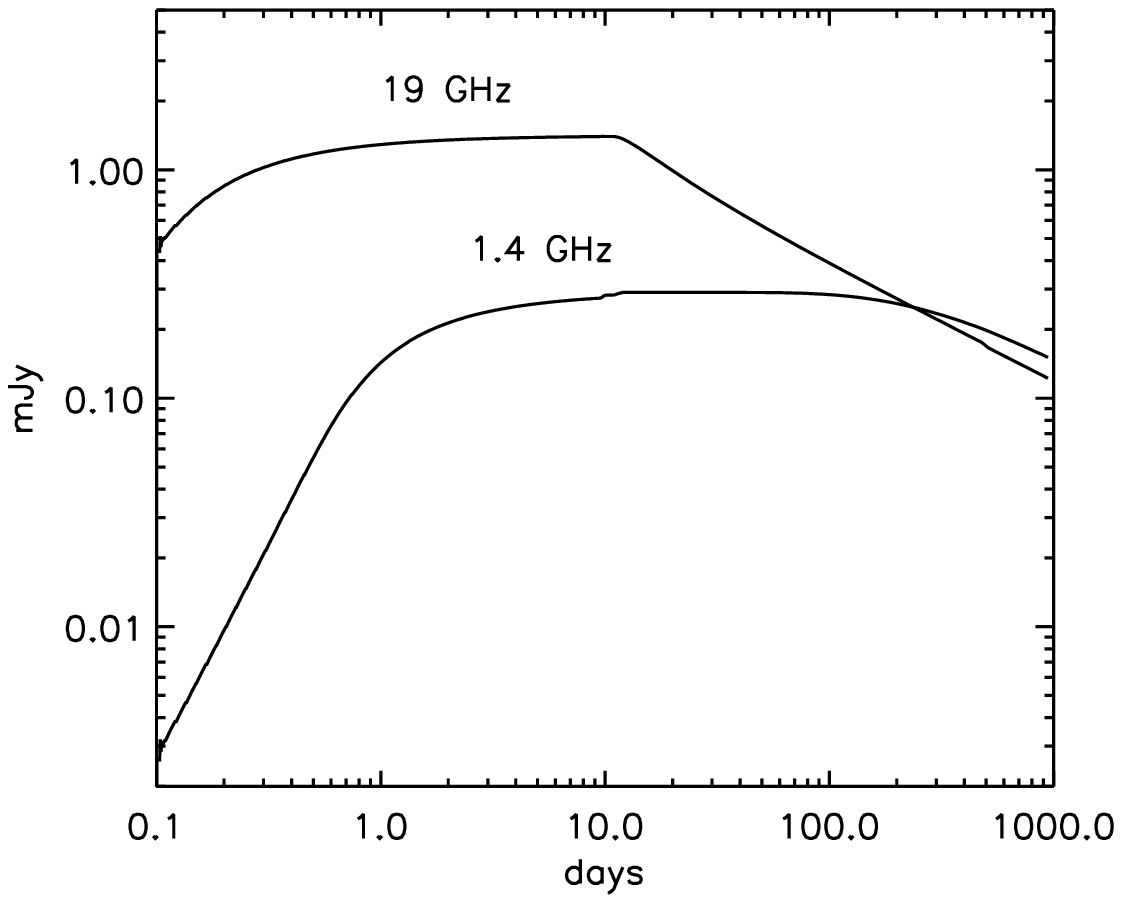} 
\caption{Radio light curves from gradual ionisation and recombination of the red giant wind, for the parameters of Run 2.}
\label{fig_radio}
\end{center}
\end{figure}

\begin{figure}[!t]
\begin{center}
\includegraphics[width= \columnwidth]{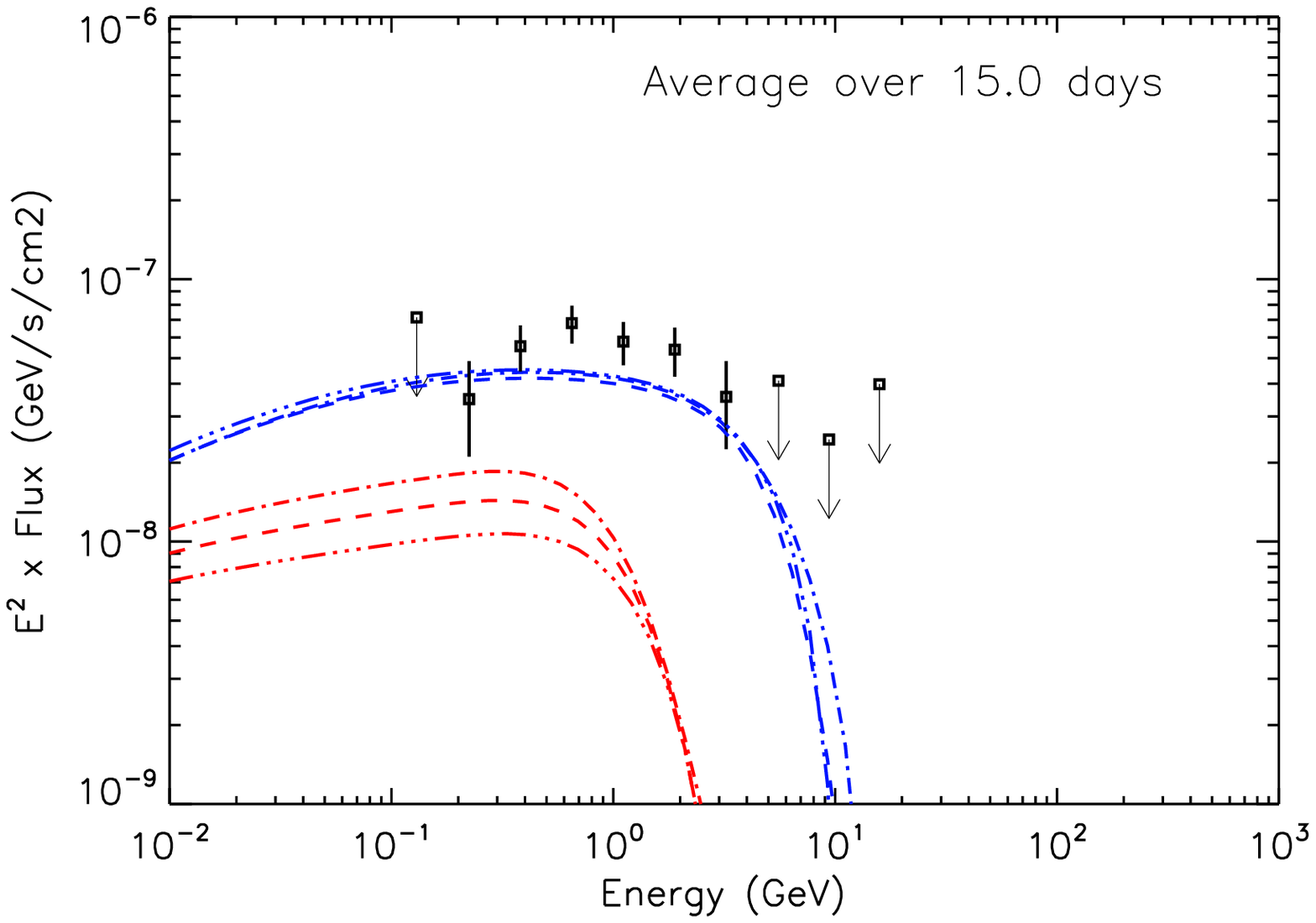}
\includegraphics[width= \columnwidth]{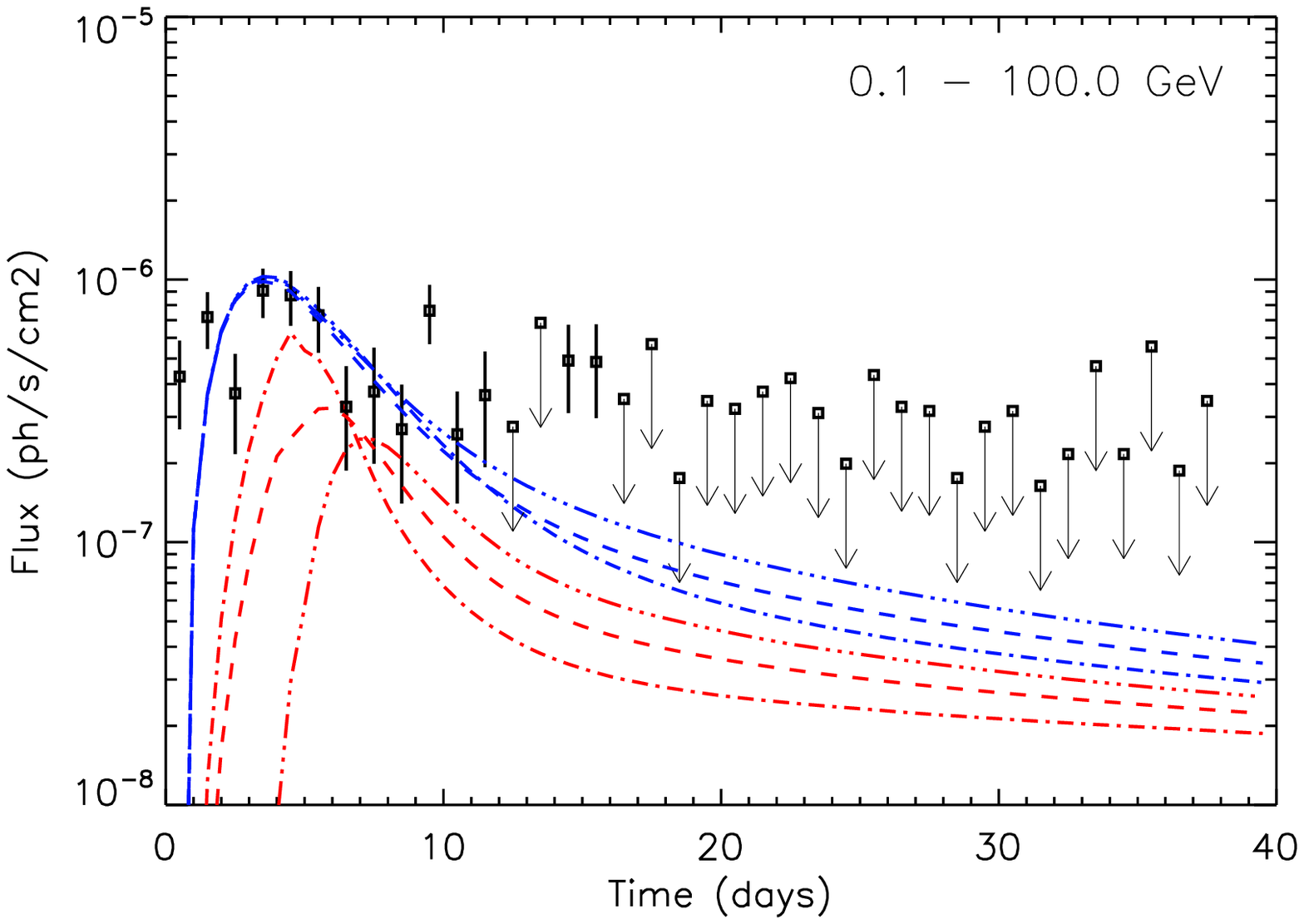}
\caption{Effect of anisotropic inverse-Compton scattering for Run 3, the scenario of a shock propagating in a wind and density enhancement around the WD. The red and blue sets of curves correspond to the scattering on the red giant and nova photons, respectively. The dot-dashed and triple-dot-dashed curves correspond to superior and inferior conjunctions, respectively, for the system viewed edge-on. The dashed curves are the isotropic case, shown for reference.}
\label{fig_emission_anisoIC}
\end{center}
\end{figure}

\indent The synchrotron emission is computed for each shock element from the upstream magnetic field conditions and the combined electron population of the acceleration and cooling zones. The calculation includes the synchrotron self-absorption and free-free absorption in the shocked layer, but no absorption from the rest of the ambient material. In the case of Run 3, the synchrotron emission rises rapidly to reach at day 4-5 a maximum of about 100\,mJy at 3\,GHz (50\,mJy at 30\,GHz), followed by a drop down to 5\,mJy at 3\,GHz at day 40 (1\,mJy at 30\,GHz). The time period of our synchrotron emission peak was not covered by radio observations, which started at day 14, but our modelled synchrotron emission past that date does not fit at all the radio data. Especially at high frequencies, the measured flux densities at day 40 and beyond largely exceeds our predictions, by a factor of about 50 for the emission at 30\,GHz. Actually, the observed flux densities increase with frequency, in contrast to the typical $F(\nu) \propto \nu^{-0.5}$ of our modelled synchrotron emission.\\
\indent Both the flux mismatch and incorrect spectral distribution suggest that the radio emission detected after the V407 Cyg outburst is not synchrotron emission. We conjecture instead that ionisation of the RG wind by the powerful UV emission during the nova outburst \citep{Shore:2011} gives rise to thermal radio emission with the required spectral index \citep{Wright:1975}. The Str\"omgren sphere increases its radius with time as radiation propagates outwards, enabling Bremsstrahlung emission. Assuming the ionisation equilibrium is instantaneous locally, this is to zeroth order equivalent to increasing the outer radius up to which Bremsstrahlung emission is integrated in \citet{Wright:1975}. For a given radio frequency $\nu$, this produces an initial $t^2$ rise of the emission, gradually flattening once the outer radius has moved beyond the characteristic radius $R(\nu)$ at this frequency (defined by $\tau_{\rm Brem}(\nu)=0.25$). The emission is then constant in time with a $F_\nu\propto \nu^{0.6}$ spectrum, as observed, until the ionisation flux drops and recombination starts. Recombination occurs on a time scale
\begin{equation}
\tau_{\rm recomb}\approx \frac{1}{n\alpha}\approx \frac{4\pi R^2 V_{\mathrm{w}} \mu m_p}{\dot{M}_{\mathrm{RG}} \alpha}
\end{equation}
where $n$ is the wind density and $\alpha$ is the recombination coefficient. Assuming ionisation stops abruptly, recombination proceeds inside-out, turning off Bremsstrahlung emission in the inner regions. According to the equation above the inner radius of the emitting wind evolves as $R_{\rm in}\propto t^{1/2}$. It can be verified that the intensity at a given frequency $\nu$ drops as $t^{-1/2}$ once the inner radius has moved beyond $R(\nu)$. The spectrum is flattish (optically thin) once decay has started and the plateau in the light curve lasts longer for lower frequencies, both of which are observed. Figure \ref{fig_radio} shows the expected light curve for two frequencies, using the parameters of Run 2. The fluxes are within an order-of-magnitude of the observations \citep{Chomiuk:2012}. Hence, this simple model suffices to capture the main features of the radio light curves. However, the predicted rise is certain to be affected by light propagation effects. The geometry, the detailed ionisation flux, the impact of the shock progression are also likely to affect the detailed evolution. These have been taken into account in the complex model presented by \citet{Chomiuk:2012}.

% Discussion
\section{Discussion}
\label{discu}

\indent In this section, we summarise the main results obtained on \object{V407 Cyg} and discuss their implications in terms of particle acceleration as well as possible alternatives that we could not explore. We then consider another instance of the phenomenon, the outburst of the \object{RS Oph} system in 2006. Last, we estimate the contribution of the whole population of such novae to Galactic cosmic rays, and the opportunities for future detections of their outbursts in $\gamma$-rays.

% Lessons learnt from V407 Cyg
\subsection{Lessons learnt from V407 Cyg}
\label{discu_v407}

\indent We aimed at modelling the {\em Fermi}/LAT observations of the \object{V407 Cyg} outburst with the nova shock accelerating particles at a moderate efficiency compatible with the test particle approximation. Similarly to cosmic-ray acceleration in SNRs, it seems that thermal X-ray emission can be a serious constraint on the non-thermal processes \citep{Ellison:2010}. The thermal X-ray emission measured by {\em Swift}/XRT imposes a relatively low density for the close environment of the binary, which limits the proton radiation efficiency and the number of particles that can be accelerated. In these conditions, the emission is mostly of leptonic origin and arises primarily from inverse-Compton scattering in the nova light. The observed level of $\gamma$-ray emission can be reproduced with a non-thermal efficiency of the shock of 10\% after two weeks and an electron-to-proton ratio at injection of 6\%. The event produced protons (respectively electrons) with energies up to $\simeq$300\,GeV (respectively $\simeq$20\,GeV), for a total non-thermal energy of 10$^{43}$\,erg. The {\em Fermi}/LAT spectrum requires shock acceleration close to the Bohm limit in the upstream equipartition magnetic field. In order to match the observed light curve, in particular the early rise and rapid drop, a density enhancement on top of the stellar wind is required in the vicinity of the WD. Such a structure very likely results from accretion and orbital motion and the need for it is corroborated by hydrodynamical simulations and studies of the thermal emission time profile. We found that the light curve and spectrum are only mildly affected by the line-of-sight orientation when realistic anisotropic inverse-Compton scattering is taken into account.\\
\indent Overall, the \object{V407 Cyg} eruption can be understood from the same principles that are invoked for particle acceleration in SNRs. An interesting point is that strong magnetic field amplification is not required, although injection of turbulence at the relevant spatial scales for Bohm diffusion of $\sim$1-100\,GeV particles is probably needed. That difference may arise from the far denser upstream medium in \object{V407 Cyg}, but whether such a diffusion regime can effectively develop given the shock physical conditions would deserve a dedicated study.\\
\indent A potential issue for the scenario described above is the weak slowing-down of the shock and ejecta. At day 40, shock velocities are everywhere above 2000\kms in our model, while line observations suggest that a substantial fraction of the ejected mass was slowed down to below 1000\kms\ at day 50. The reconstruction of the shock kinematics from emission line signatures is not an easy task, and it may just be that we are comparing the velocities of distinct emitting regions. Optical spectra of the \object{V407 Cyg} nova reported in \citet{Shore:2011} show a diversity of line shapes and evolutions, which may be indicative of a complex underlying flow pattern with a large variety of physical conditions. As an example, the maximum radial velocities for the H$\alpha$ and He II transitions differ by almost an order of magnitude 40 days after outburst, from about 3000 to 300\kms\ for the blueshifted side of the lines. On top of that, part of the discrepancy may simply result from the rough prescriptions adopted for the shock evolution in our model, which basically neglected all the downstream hydrodynamics. Radiative hydrodynamics simulations would be needed to really explore that aspect of the problem. Nevertheless, we investigated the effect of a more efficient shock slowing-down by forcing the velocity to drop faster so that it reaches below 1000\kms\ after day 40. The most interesting consequence is that the non-thermal efficiency does not increase at late times and is frozen at the value it has at about day 10. Apart from that, the spectrum and light curve are almost unaffected.\\
\indent Last, we want to emphasise that the test particle approximation led to dominantly leptonic emission and a high electron-to-proton ratio at injection, for the global non-thermal efficiency had to remain relatively low. We cannot exclude, however, the possibility of a comparable, if not dominant, hadronic contribution to the emission, but that would require very efficient proton acceleration and our model is not suited to the study of cosmic-ray dominated shocks (in addition, the contribution of secondary electrons/positrons from charged mesons would need to be taken into account). For the same binary and nova parameters as in Run 3, the {\em Fermi}/LAT data can be reproduced with $\eta_{\mathrm{inj,p}}=2 \times 10^{-2}$, $\eta_{\mathrm{inj,e}}=2 \times 10^{-4}$, and $\zeta=5$, but then some parts of the blast wave would convert $>50$\% of their initial kinetic energy into accelerated particles. Although we cannot rule out that option, we note that the study of another nova in a symbiotic system, \object{RS Oph}, concluded that particle acceleration should have proceeded with a relatively moderate efficiency (see below).

% Another nova outburst, RS Oph
\subsection{Another nova outburst, RS Oph}
\label{discu_rsoph}

\indent There are many similarities between \object{V407 Cyg} and \object{RS Oph} \citep{Abdo:2010c,Munari:2011}, although a thorough analysis of emission line signatures revealed interesting differences between the \object{V407 Cyg} 2010 event and the \object{RS Oph} 2006 event \citep{Shore:2011}. \object{RS Oph} is a symbiotic binary consisting of a massive WD and a RG, separated by about 2\,AU and orbiting with a period of 455\,days \citep{Fekel:2000}. The system is located at a distance of 1.6\,kpc and was observed in eruption on six occasions in 1898, 1933, 1958, 1967, 1985, and 2006. The last event in particular was covered in X-rays by {\em Swift}/XRT, but no $\gamma$-ray instrument was available at that time to observe in the $>$100\,MeV range.\\
\indent We used our model to estimate the high-energy emission from the 2006 eruption of \object{RS Oph}. The system parameters are similar to those adopted in \citet{Walder:2008} and \citet{Tatischeff:2007}: binary separation $d_{\mathrm{orb}}$= 2\,AU, RG radius $R_{\mathrm{RG}}$= 150\rsol\ and mass-loss rate $\dot{M}_{\mathrm{RG}}$= 10$^{-7}$\wunit, ejecta mass $M_{\mathrm{ej}}$= 10$^{-6}$\msol\ and initial velocity $V_{\mathrm{ej}}$= 4000\kms (other parameters being similar to the base case for \object{V407 Cyg}). With these assumptions, the wind density at the RG surface is about $10^{9}$\,cm$^{-3}$. The hydrodynamical simulations by \citet{Walder:2008} showed that the WD accretion and orbital motion result in a density enhancement of about $10^{11}$\,cm$^{-3}$ at the WD position, with a typical scale length of 1\,AU (for comparison, we inferred for \object{V407 Cyg} a density enhancement at the WD of the order of the density at the RG surface with a scale length of 10\,AU).\\
\indent In the case of particle acceleration occurring at the Bohm limit in the upstream equipartition field, the 2006 eruption of \object{RS Oph} produced protons reaching up to almost 1\,TeV and electrons reaching about 30\,GeV. This is quite similar to what was obtained for \object{V407 Cyg} in the same acceleration conditions (and confirms once again the robustness of the maximum particle energy to changes of the binary and nova parameters). Due to the massive accumulation of gas close to the WD, the predicted $\gamma$-ray emission from \object{RS Oph} is dominantly of hadronic origin, even for electron-to-proton at injection up to 10\%. For a global non-thermal efficiency of $\simeq$10\%, the $>100$\,MeV flux rises up to 10$^{-4}$\funit\ within a day, but subsequently drops by more than three orders of magnitude over the following 3-4 days. This $\gamma$-ray flash actually is due to the strongly peaked matter distribution in \object{RS Oph}, where a relatively high density of target particles for pion production is found only in the immediate vicinity of the WD. Shock deceleration takes place very rapidly within the first 2 days and the level of thermal X-ray emission after 3 weeks is a few times $10^{35}$\,erg\,s$^{-1}$ in agreement with \citet{Sokoloski:2006}. The different thermal X-ray signatures at early times in the \object{RS Oph} 2006 event, which displayed powerful emission within a few days after outburst, and in the \object{V407 Cyg} 2010 event, which exhibited a late rise to a more modest emission level, very likely arise from the matter distribution in the innermost regions of the binary system (namely the density enhancement on top of the stellar wind).\\
\indent Particle acceleration in \object{RS Oph} was studied by \citet{Tatischeff:2007} and constrained through its influence on the shock structure and X-ray emission. The authors estimated that protons could be accelerated up to a few TeV in a few days but are prevented to go beyond by particle escape upstream of the shock. Although both approaches are different and we did not consider upstream escape, this is consistent with our result for very similar assumptions about the particle diffusion regime. Then, using a semianalytic model of nonlinear DSA, the authors found that the post-shock temperature evolution can be accounted for if the shock converts up to about 60\% of the instantaneous energy flux into non-thermal particles and is efficiently cooled through the escape of accelerated particles. Yet, the implied particle acceleration efficiency remains moderate, as in the scenario presented above for \object{V407 Cyg}.\\
\indent Finally, we note that there are more objects likely to be other examples of a fast nova explosion in a symbiotic system. The dramatic X-ray, optical and radio 1998 flare of \object{CI Cam} may be the result of a classical nova eruption in the dense circumstellar medium of a Be star \citep{Filippova:2008}. More recently, the transient X-ray source \object{MAXI J0158-744} detected in the Small Magellanic Cloud was interpreted as a nova in a Be-WD binary \citep{Li:2012}.

% The population of novae in symbiotic systems
\subsection{The population of novae in symbiotic systems}
\label{discu_pop}

\indent From the above elements, we can estimate the contribution of novae in symbiotic systems to the production of Galactic cosmic rays. Their occurrence rate is in the range 1-10\,yr$^{-1}$, but not all systems are likely to be equally efficient \citep{Lu:2011}. Assuming that the 2006 event of \object{RS Oph} and the 2010 event of \object{V407 Cyg} are representative, novae in symbiotic systems would release of the order of 10$^{43}$\,erg in the form of $<$1\,TeV protons. For comparison, SNRs are 100 times less frequent but are believed to release at least 10$^{49}$\,erg of $<$1\,TeV protons. The population of objects exemplified by \object{V407 Cyg} and \object{RS Oph} is therefore not expected to be a significant contributor to Galactic cosmic rays.\\
\indent In terms of perspective for future $\gamma$-ray observations of novae in symbiotic systems, our work suggests that the best candidates would be binaries with relatively short period of order 1-10\,yr because of the high peak luminosities they can reach if they effectively develop a significant density enhancement around the WD (this, however, may be counterbalanced by a very rapid emission decrease for the closest systems). Optimistically taking our model for the 2006 \object{RS Oph} nova as a reference (because it is brighter in $\gamma$-rays), such a source could probably be detected by the {\em Fermi}/LAT up to distances of $\simeq$10\,kpc. Within that horizon, the occurrence rate of all novae in symbiotic systems, irrespective of their orbital period, would be at most a couple of events per year (assuming a uniform distribution of sources and a Galactic disk of 20\,kpc radius). Hence, no more than a few events like \object{V407 Cyg} can reasonably be expected in the {\em Fermi}/LAT remaining 5-10 years of operations. In any case, our analysis suggests that novae in symbiotic systems are unlikely to be a class of TeV-emitters \citep{Aliu:2012}.\\
\indent As we finished this work, the {\em Fermi}/LAT collaboration reported evidence for $\gamma$-ray emission associated with two classical novae: Nova Sco 2012 \citep{Cheung:2012} and Nova Mon 2012 \citep{Cheung:2012a}. In these cases, the companion is suspected to be a low-mass main-sequence star in a tight orbit with a typical period of order of an hour, but the nature of the progenitors is still unknown. Expansion of the burning WD envelope can be accompanied by a supersonic shock as in \object{V407 Cyg}, so particle acceleration is also possible. The stellar radiation field from the companion is very weak but inverse-Compton emission could still occur on photons from the nova, whose contribution dominates the $\gamma$-ray light curve in our model for \object{V407 Cyg}. However, the stellar wind is too tenuous to slow down the nova ejecta, which rapidly expands to engulf the whole system. Hence, little swept-up material can be accelerated and significant $\gamma$-ray emission is puzzling if the environment simply consists of the stellar wind. One possibility is that shocks in classical novae interact with the cold, massive circumbinary discs that are thought to surround cataclysmic variables up to distances of several AU \citep{Spruit:2001,Dubus:2004}, in a way that is reminiscent of the setup of Run 3 that we presented here for \object{V407 Cyg}. A different possibility is that particle acceleration occurs at the temporary bow shock or convergent flow that forms behind the companion star as the ejecta expands beyond the binary orbit. Alternatively, it may be that in some novae the ejecta are expelled with some velocity gradient, such that internal shocks may develop and accelerate particles from the ejecta itself rather than from the circumstellar environment \citep{Lloyd:1992,OBrien:1994}. The rate for classical novae in the Galaxy is about 10 times higher than for symbiotic novae \citep{Lu:2006}, so future {\em Fermi}/LAT observations should be able to shed more light on this unexpected phenomenon.

\begin{acknowledgement}
The authors acknowledge support from the CNES, and from the European Community via contract ERC-StG-200911. We thank Guy Pelletier for helpful discussions on particle acceleration and Pierre Jean for useful comments. The authors also are grateful to Salvatore Orlando and Ulisse Munari for providing additional information about their respective works.
\end{acknowledgement}

\bibliographystyle{aa}
\bibliography{/Users/pierrickmartin/Documents/MyPapers/biblio/SNRmodels,/Users/pierrickmartin/Documents/MyPapers/biblio/SMC,/Users/pierrickmartin/Documents/MyPapers/biblio/CosmicRaySources,/Users/pierrickmartin/Documents/MyPapers/biblio/CosmicRayTransport,/Users/pierrickmartin/Documents/MyPapers/biblio/GalaxyObservations,/Users/pierrickmartin/Documents/MyPapers/biblio/DataAnalysis,/Users/pierrickmartin/Documents/MyPapers/biblio/Fermi,/Users/pierrickmartin/Documents/MyPapers/biblio/Books,/Users/pierrickmartin/Documents/MyPapers/biblio/SNobservations,/Users/pierrickmartin/Documents/MyPapers/biblio/26Al&60Fe,/Users/pierrickmartin/Documents/MyPapers/biblio/Positron,/Users/pierrickmartin/Documents/MyPapers/biblio/44Ti,/Users/pierrickmartin/Documents/MyPapers/biblio/Cygnus&CygOB2,/Users/pierrickmartin/Documents/MyPapers/biblio/Physics,/Users/pierrickmartin/Documents/MyPapers/biblio/V407Cyg}

\end{document}